\documentclass[prl,preprint,aps,showpacs,showkeys]{revtex4-2}
\usepackage{amsmath,amssymb,graphicx,graphics,latexsym,xcolor}
\usepackage{caption}
\usepackage{subcaption}
\usepackage{enumerate}
\usepackage{hyperref}
\usepackage{natbib}
\usepackage{silence}
\usepackage{float}
\usepackage{calligra} 
\DeclareMathAlphabet{\mathcalligra}{T1}{calligra}{m}{n} \DeclareFontShape{T1}{calligra}{m}{n}{<->s*[2.2]callig15}{}

\newcommand{\A}{\mathcal{A}}
\newcommand{\rr}{\ensuremath{\mathcalligra{r}}}

\DeclareMathOperator{\Hl}{H\ell}
\DeclareMathOperator{\Hf}{Hf}
\DeclareMathOperator{\Hp}{Hp}
\newcommand{\K}{\mathop{\raisebox{-5pt}{\huge K}}}

\begin{document}
\title{\Large \bf Characterizing a class of accelerating wormholes with periodic potential}

\author{Soham Chatterjee} \email{avich1010@gmail.com}
\author{Sagnik Roy} \email{roysagnik134@gmail.com}
\author{Ratna Koley} \email{ratna.physics@presiuniv.ac.in}
\affiliation{Department  of  Physics,  Presidency  University,  86/1 College Street, Kolkata  700073,  India}

\begin{abstract}
The newly discovered Wormhole C--metric is a solution of Einstein's field equation coupled with a phantom scalar field which describes the accelerated wormholes. In the zero acceleration limit the solution reduces to an asymptotically flat wormhole. 
For certain range of parameter space this solution doesn't possess any horizon, thus making it a viable candidate of wormhole. 
To completely unveil this property we have studied the topological properties of this spacetime and shown that the throat is marginally connected. In the aforementioned range of parameters, the spacetime doesn't posses any photon orbit confirming the absence of shadow. We further analysed the stability of this spacetime under scalar perturbation. Under the usual boundary conditions (outgoing waves at both spatial infinities) there exists a continuous spectra. On the contrary one may achieve the quantization of the modes by exploiting a different but physically intuitive boundary condition. The lowest lying mode behaves as normal mode, and the imaginary part comes into play for the modes corresponding to first overtone number $(n=1)$ marking the onset of quasi-nomral modes for all azimuthal quantum number, $L$. We have also argued that the spacetime has a tendency to hold the excitation in it due to the external perturbation, rather than a fast de-excitation.  
\end{abstract}


\maketitle
\newpage
\section{Introduction}

The C--metric, a family of vacuum solutions of the Einstein equation, characterizes a pair of causally disconnected black holes with a uniform acceleration \cite{Kinnersley,Bonnor1983,Ads_accl_bh,pair_accl_bh_ads,uniform_accl_bh}. The boost-rotation symmetry of the solution brings it under the general class of Plebański--Demiański spacetimes (and hence Petrov D) and the Weyl spacetimes. The acceleration of each black hole arises either from  a conical deficit angle along the symmetry axis, akin to a semi-infinite cosmic string, or from a strut with negative tension developing between two black holes \cite{Griffiths_c_metric}. In the linear 
approximation of the acceleration, the C--metric emerges as a perturbation of the Schwarzschild black hole with a distributive stringy source, or as a nonlinear superposition of the Schwarzschild and Rindler spacetimes.
Consequently, the geodesics in a C--metric represent the trajectories of test particles under the gravitational field of a black hole subject to a uniform constant force -- commonly known as the \textit{gravitational Stark effect} \cite{Grav_stark_effect}. Extensive studies have delved into the causal structures, as well as the physical and geodesic properties \cite{Lim_2014,Lim_2021,Perlick_2021,coacc_c_geodesic,spinning_c_geodesic}, of the black hole C--metric. These investigations remain pivotal within the Physics community, as not only it is useful to model accelerated black holes, but also it can shed light on the binary black hole mergers, causing gravitational radiation \cite{accl_bh_grav_rad}. \\

Another interesting class of solution of the Einstein field equation is the wormhole spacetime. The solution describes a topologically non-trivial and non-singular geometry which can act as a tunnel between different universes or different parts of the same universe. Traversable wormhole was conceived by Morris, Thorne and Yurtsever \cite{Morris_Throne_wormhole,MTY} where the geometry is supported by the energy condition violating exotic matter. Many such traversable wormhole solutions have been extensively studied over the last few decades \cite{trw1,trw2,trw3,trw4,trw5,trw6,trw7,trw8,trw9,trw10,trw11}. 
Recently, Nozawa and Torii have proposed \cite{Wormhole_C-metric} the accelerated version of wormhole by the suitable analytic continuation of one parameter family of the $\mathcal{N}=2$ gauged supergravity -- termed as the wormhole C--metric. This is a solution of Einstein's gravity coupled with a phantom scalar field, expressed in terms of a superpotential. The potential exhibits multiple $AdS$ extrema, with local minima corresponding to critical points of the superpotential. Through a flipping transformation, the metric undergoes a transition to a different family of C--metrics and upon approaching the zero acceleration limit, each solution converges towards two wormhole solutions. Within a natural coordinate framework, the C--metric solution approaches $AdS$ at the origin of the potential. Through maximal extension, the solution seamlessly connects to the other side of the universe, where the scalar field evolves towards a distinct critical point of the potential. An interesting characteristic of these C--metrics is their ability to avoid the conical singularities typically required for inducing acceleration  under appropriate parameter choices. Instead, the acceleration in these scenarios is solely facilitated by the phantom character associated to the scalar field in contrast to the black hole C--metric.\\

In this article, the geodesics and quasinormal behaviour of the wormhole C--metric have been explored extensively which will be useful to find the possible observational imprints. At first we study the geodesics and shadow profile in this geometry for a better understanding of the spacetime. The shadow of an object is quite intriguing as it is closely related to the behaviour of geodesic motion of photon in the vicinity of the object \cite{Perlick_shadow_review,wormhole_shadow,Lorentzian_wormhole_shadow}. This direction of research has been resurrected with the observation of black hole M87 shadow by EHT in 2019, followed by the detection of Sagittarius A* in 2022 \cite{EHT1,EHT2,EHT3,EHT4,EHT5}. 
%
Various wormhole shadows have been studied over the past decade with the aim of distinguishing them from the shadows of ordinary black holes, which could potentially aid in probing these exotic objects. For instance, in the case of a traversable wormhole, the central region of the shadow is directly illuminated by low angular momentum photons coming from the other side of the wormhole. Additionally, images of reflection-asymmetric wormholes \cite{refas1, refas2} may contain a region where photons emitted on one side of the wormhole travel to the other side and are reflected back to their side of origin. Conversely, if the wormhole is rotating (see, e.g., \cite{trw6}), then for small rotation and small throat size, it can closely mimic a Kerr black hole with the same rotation parameter and mass. However, as the spin or throat size increases, the shadow of the Kerr black hole starts to deform more than that of the rotating wormhole shadow. Using this distinctive feature, Rahaman et al. \cite{rahaman} constrained the wormhole parameters from the M87* shadow. A similar feature was also observed in \cite{Shaikh}. There is extensive literature on these topics, and readers may refer to \cite{whsh1,whsh2,whsh3,whsh4,whsh5,whsh6,whsh7,whsh8,whsh9,whsh10,whsh11,whsh12,whsh13,whsh14,whsh15,Perlick_shadow_review} for further details.  In the spacetime under consideration the null energy condition has been shown to be violated not only near the throat but across the entire spacetime for the associated parameter space. Surprisingly the spacetime does not possess any photon orbit resulting in absence of a shadow. This is a very interesting feature in the sense that previously no spacetime solution was found - to the best of our knowledge - for which electromagnetic shadow doesn't form.
The detection of this kind of spacetime may reveal interesting observational aspects as well as a probe for potential verification of the underlying quantum gravity theory. Astronomical objects are primarily detected through two main methods: electromagnetic observations which involve studying phenomena such as accretion or observation of the shadow profile cast by the spacetime and the detection of gravitational waves. \\

It is well established that the merger of binary black holes or compact objects results in emission of gravitational waves. There are mainly three phases in this process: inspiral, merger and ringdown. During these phases enormous amount of energy is released in the form of gravitational waves. In 2016, first direct detection of gravitational wave was made by LIGO and Virgo \cite{Ligo_GW}. Generally the astrophysical objects are not isolated and they are always subject to some kind of perturbations. The stability of a given spacetime against these perturbations is an important question. The black holes exhibit damped oscillations under perturbation due to the radiation of gravitational waves \cite{Berti_qnm,Konoplya_qnm} and these are described by the quasi-normal modes (QNMs). The QNM waveform produces a ringdown signal until the two objects are settled to a stable final object and these late-time tails carry important information about the final object. The QNM of accelerating black hole was calculated by Destounis et. al. \cite{Accelerating_BH_QNM}. In this work we study the QNMs in the given spacetime and the effect of boundary conditions on the QNMs in greater detail. Interestingly we have found that although the spacetime is stable it holds the excitation due to the perturbation rather than getting de-excited and coming to the ground state. These important features demand attention both 
from theoretical and observational perspectives. \\

The work is organised in the following way. In Section \ref{section:wormholeCmetric}, the physical properties and the causal structures of the wormhole C--metric has been analysed. The null geodesic has been studied for a special type of mass parameter. Section \ref{section:geodesic} deals with the geodesic properties and the absence of shadow profile in the spacetime. In Section \ref{sec:ScalarPerturbation}, we first discussed the conformal structure of the spacetime. Then for the standard choice of boundary conditions (i.e. the waves are outgoing at the special infinities), the scalar modes of the spacetime have been calculated and it is  found to be continuous. Then for a new choice of the boundary conditions the modes have been shown to be quantized. Physical 
interpretation of the results have been given. Finally we concluded in Section \ref{sec:conclusion}. In Appendix \ref{appendix:notations} the notations, conventions and the symbols used throughout the article have been mentioned.

\section{Wormhole C--metric spacetime}\label{section:wormholeCmetric}

The $4$-D Einstein gravity minimally coupled with a real scalar field can be described by the action 
\begin{equation}
    \mathcal{S} = \dfrac{1}{2} \displaystyle\int \boldsymbol{\sqrt{-\Tilde{g}} d^4x} \left[R - \epsilon_1(\partial\Phi)^2 - 2\mathcal{V}(\Phi)\right] \label{action}
\end{equation}
where $\epsilon_1=+1$ corresponds to the ordinary scalar field, whereas $\epsilon_1=-1$ represents the phantom scalar field mediating the repulsive force. 
Our aim is to explore the consequences of the phantom property of the scalar field throughout the paper. The potential of the phantom scalar field is denoted by \(\mathcal{V}(\Phi)\) that can be written as
\begin{equation}
    \mathcal{V}(\Phi) = -g^2\left(2 + \cos{\sqrt{2}\Phi}\right) \label{scalar_field_potential}
\end{equation}
where $g$ acts as an overall scale factor of the potential that can be chosen non-negative without any loss of generality, which we will continue to follow for the rest of the paper. As a special case, the potential can be set to be zero by letting $g=0$. Here only the periodic and phantom nature of the potential has been considered for the sake of convenience, but for more general case, one can refer \cite{Wormhole_C-metric}.

Now for the system \eqref{action} and \eqref{scalar_field_potential}, the C-metric solution takes the form \cite{Wormhole_C-metric}
\begin{eqnarray}\label{C-metric_solutions_old_coordinates}
    ds^2 &=& \dfrac{1}{(1+Arx)^2} \left[V(x) \left\{-\Delta_r(r) d\tau^2 + \dfrac{dr^2}{\Delta_r(r)}\right\} + r^2 V_r(r) \left\{\dfrac{dx^2}{\Delta_x(x)} + \Delta_x(x) d\varphi^2\right\}\right] \\
    \Phi &=& \sqrt{2} \left[\tan{^{-1}\left(\dfrac{r}{m}\right)} - \tan{^{-1}\left(Amx\right)} - \dfrac{\pi}{2}\right]
\end{eqnarray}
Clearly for $m=0$, the scalar field, $\Phi$ becomes trivial and thus let us refrain from considering this. In Eq. \eqref{C-metric_solutions_old_coordinates}, $A$ denotes the acceleration of the spacetime and a dimensionless acceleration parameter can be defined out of this: $\A=Am$. Further defining a dimensionless radial coordinate $\rr=r/m$ and a dimensionless scale factor $g$ by replacing $gm$ and considering the global topology of $x-\varphi$ to be $\mathbb{S}^2$, the solution reduces as
\begin{eqnarray}\label{C-metric_solutions}
    ds^2 &=& \dfrac{1}{(1+\A\rr x)^2} \left[V(x) \left\{-\Delta_\rr(\rr) d\tau^2 + \dfrac{d\rr^2}{\Delta_\rr(\rr)}\right\} + \rr^2 V_\rr(\rr) \left\{\dfrac{dx^2}{\Delta_x(x)} + \Delta_x(x) d\varphi^2\right\}\right] \label{C-metric} \\
    \Phi &=& \sqrt{2} \tan{^{-1}\dfrac{1+\A\rr x}{\A x-\rr}} \label{capital_phi}
\end{eqnarray}
where $V(x)$, $\Delta_\rr(\rr)$, $V_\rr(\rr)$, $V_\rr(\rr)$ must be positive in order to maintain the Lorentzian signature of the metric and thus can be defined as
\begin{eqnarray}
    V(x) &=& 1+\A^2x^2, ~~ \Delta_\rr(\rr) = 1+g^2+\left(g^2-\A^2\right)\rr^2, ~~ V_\rr(\rr) = 1+\dfrac{1}{\rr^2}, ~~ \Delta_x(x) = 1-x^2
\end{eqnarray}
As mentioned in \cite{Wormhole_C-metric}, $x$ represents the direction cosine on $\mathbb{S}^2$ and hence can be taken to be proportional to $\cos{\theta}$. But unlike the traditional C--metrics, the two dimensional surface spanned by $x-\varphi$ of this spacetime, being free from conical singularity, $x$ can be considered as $\cos{\theta}$ only. It is also worth mentioning that one do not need distributional sources to maintain the acceleration of wormholes, which stands in stark contrast to the vacuum case.

\subsection{Infinity}
This spacetime possesses two killing vectors: $\partial_\tau$ and $\partial_x$, which correspond to the two conserved quantities, energy $(E)$ and angular momentum $(L)$ respectively. The affine parameter along a null curve with $(\tau,x,\phi) =$ constant can therefore be written as \cite{affine_parameter_calculation_Visser}
\begin{eqnarray}
    \lambda \sim \left|\dfrac{\A\rr\left(1+\A^2x^2\right)}{E(1+\A\rr x)}\right|^{1/2}
\end{eqnarray}
which clearly indicates that it takes an infinite amount of ``time'' to reach any point on the surface $\rr=-\frac{1}{\A x}$, whereas $\rr\to\infty$ surface can be reached with a finite affine parameter. Hence unlike the standard Morris--Thorne wormhole \cite{Morris_Throne_wormhole} (or Lorentzian wormholes \cite{Visser_Lorentzian_wormholes}), $\rr\to-\frac{1}{\A x}$ will be treated as \textit{infinity} for this spacetime. Obviously at the equatorial plane, the infinity lies at $\rr\to\infty$, but not in general.

\subsection{Curvature properties and energy conditions}
\begin{figure}[H]
\centering
\subfloat{\includegraphics[width=.45\linewidth]{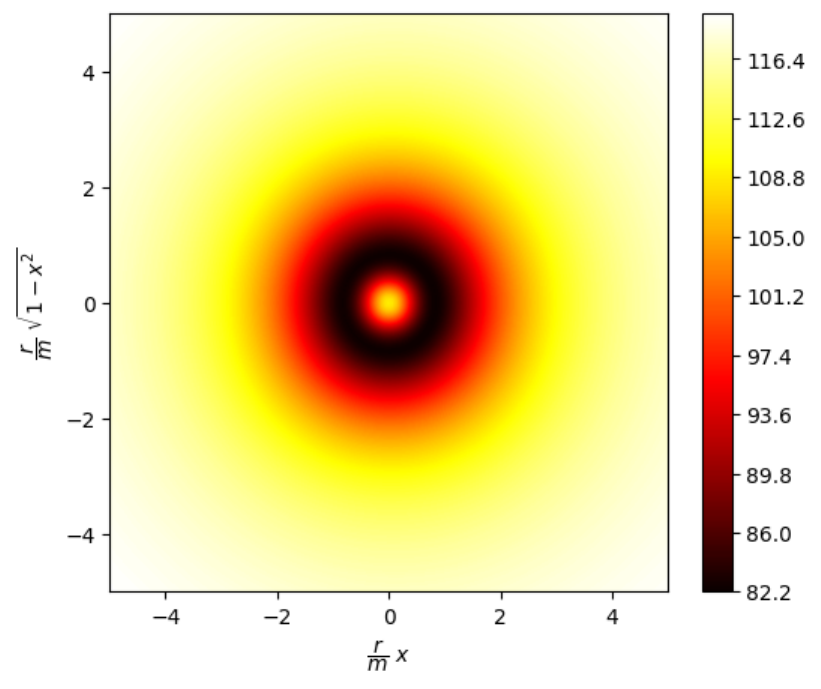}} \hspace{10mm}
\subfloat{\includegraphics[width=.45\linewidth]{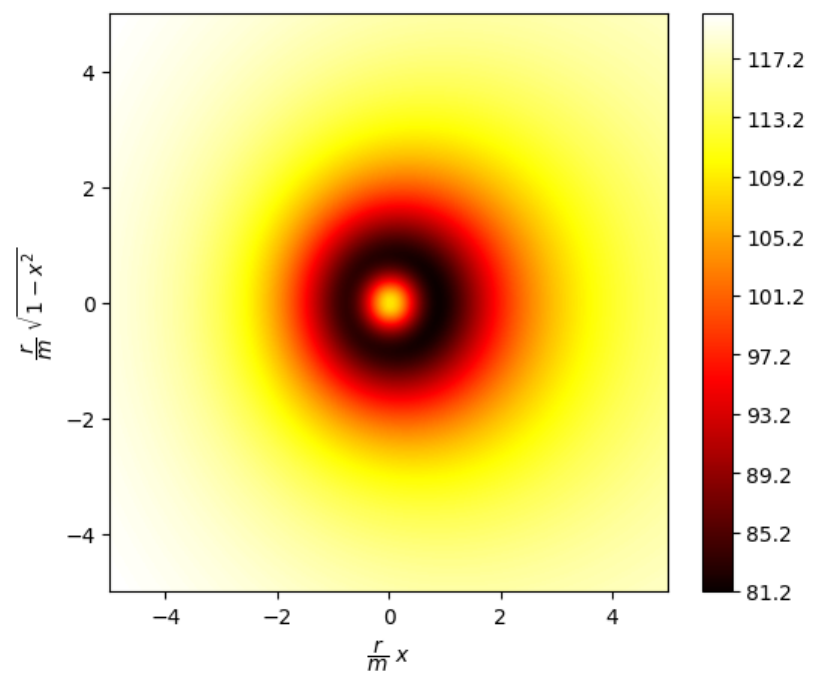}} \par
\subfloat{\includegraphics[width=.45\linewidth]{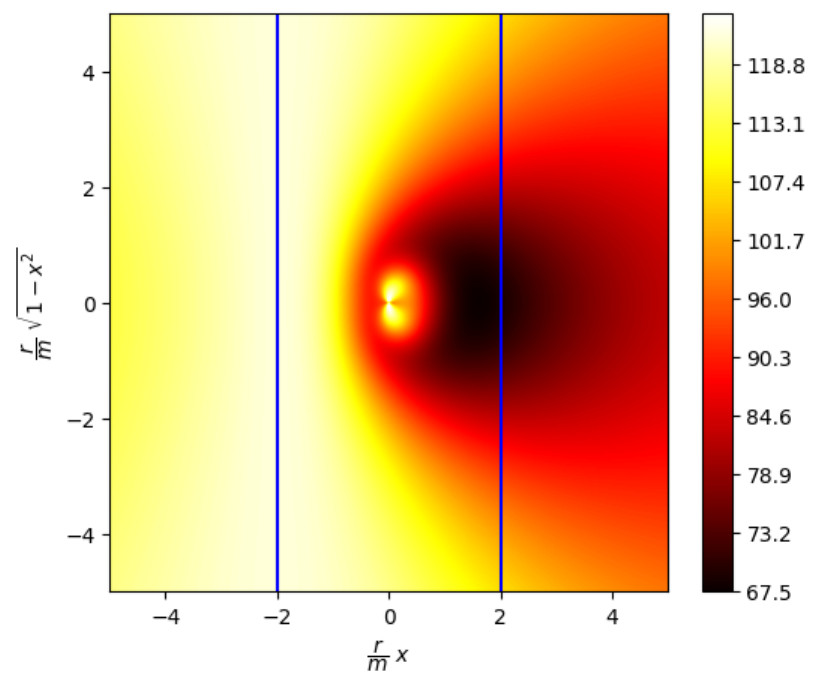}} \hspace{10mm}
\subfloat{\includegraphics[width=.45\linewidth]{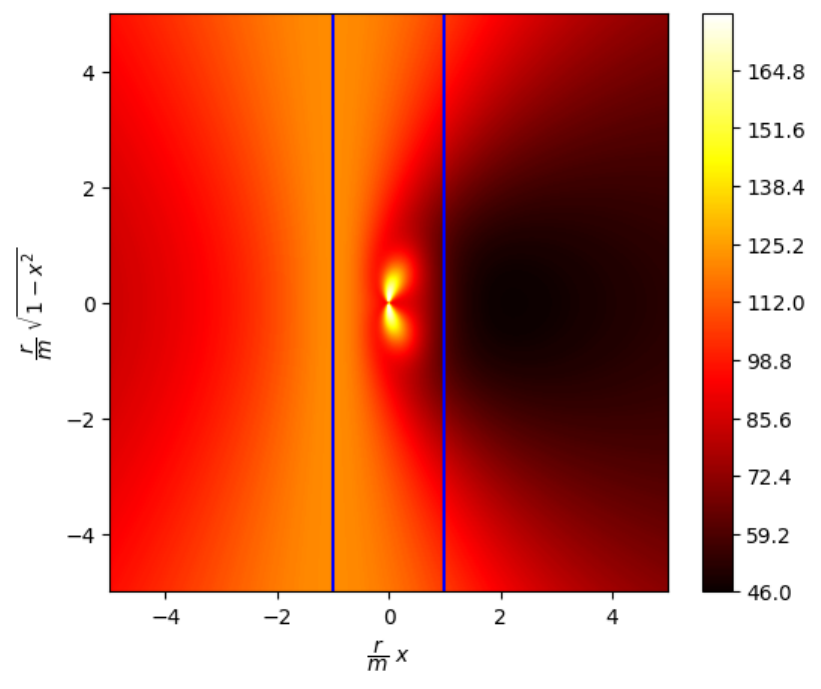}}
\caption{Variation of Kretschmann scalar for $\A=5\times10^{-3}$, $\A=5\times10^{-2}$, $\A=0.5$, $\A=1.0$ (from upper left to lower right) for a fixed value of scaling factor, $g=1.5$.}
\label{fig:Kretschmann}
\end{figure}

The existence of curvature singularities in a Lorentzian manifold is usually determined by the divergence of the Kretschmann scalar $(K = R^{\alpha\beta\gamma\delta}R_{\alpha\beta\gamma\delta})$ as this remains invariant under coordinate transformation. Therefore, if $K$ diverges in one frame of reference, it will also be divergent in all other coordinate frames, removing the ambiguity associated with the co--ordinate singularity. For the spacetime metric in Eq. \eqref{C-metric}, the analytical expression of this quadratic scalar invariant is too cumbersome. That is why instead of writing the analytical form explicitly, the variation of $K$ over $x-\varphi$ surface has been depicted pictorially in Fig. (\ref{fig:Kretschmann}).

From the figure, it is evident that the isocurvature surfaces of the spacetime exhibit a degree of symmetry for very low acceleration, which is expected since the spacetime itself becomes spherically symmetric under such conditions. As the acceleration parameter increases the assymetrical deformation of these surfaces starts on one side of the wormhole. Additionally the curvature of the spacetime does not monotonically increase or decrease with radial distance; rather there exists a plateau region (indicated by the black region in Fig. \ref{fig:Kretschmann}). The Kretschmann scalar (and hence the spacetime curvature as a whole) does not diverge anywhere in the domain, as indicated by the analytic expression of $K$. Therefore the spacetime does not exhibit any curvature singularity. Furthermore, it is noteworthy that the spacetime lacks asymptotic flatness near infinity, as observed in Fig. \ref{fig:Kretschmann}, a feature corroborated by the line element \eqref{C-metric}
Considering the future oriented null vectors \(k^\mu = (\sqrt{-g_{00}}, -\sqrt{g_{11}}, 0, 0)\) and \((\sqrt{-g_{00}},0,-\sqrt{g_{22}},0)\), the violation of null energy condition requires only the positivity of $\Delta_\rr(\rr)$ and $\Delta_x(x)$ over the entire spacetime, respectively, which is always satisfied. Therefore the energy condition is always being violated in this case, resulting in the spacetime being classified as a wormhole spacetime. Although the violation of null energy condition alone can't justify the spacetime being a wormhole, it requires rigorous topological analysis to conclude this, which has been discussed briefly in Appendix \ref{topology_wormhole}.

\subsection{Causal structure}
With the previous rescaling of coordinates, $r\to m\rr$, $A\to\A/m$ and $g\to g/m$, the $3$ parameters of the metric have reduced effectively to $2$ parameters. Now considering the fact that the line element without conformal factor admits the mirror symmetry, $\A\to-\A$ and considering $\A$ to be non--negative without losing any generality, an analytic classification of the ranges of values of $\A$ is possible: $0\le\A<g$, $\A=g$ and $g<\A<\infty$. Among these cases, $\A=g=0$ corresponds to the well-known Ellis wormhole with drain-hole parameter $=1$ and for $g=0$, it takes the form of a generalized Ellis--Bronnikov wormhole. 
However, we will consider the case where $\A > g$. Under these constraints, there will be two horizons located on both sides of the wormhole at $\rr=\pm1/\A$. For detailed discussions on the characterization of these horizons and the corresponding Penrose diagrams, one may look into the original paper of \textit{Wormhole C--metric} \cite{Wormhole_C-metric}. Now in case of the presence of these horizons, as evident from the line element \eqref{C-metric}, the outer region of the horizon is spacelike, whereas the inner part is timelike. So if we admit that the observer is in a timelike region of the spacetime, then the observer must live inside the horizon, which is bizarre. Also if it is considered that the observer is inside the horizons, then it will not get access to the spatial infinity (on the equatorial plane, spatial infinity can \emph{marginally} be accessed sitting inside the horizons though), which is non-intuitive as well. Therefore it is safe to exclude the parameter space of $\A$ and $g$, for which the spacetime admits horizons. For the case of $\A=g$, the spacetime reduces to the well known Ellis--Bronnikov wormhole spacetime by the slight redefinitions of $\tau$ and $\rr$ coordinates. The only non--trivial case left is $0\le\A<g$, which we will assume to be valid in the subsequent analysis.

For $\A<g$, it is possible to extend Visser's proposal of defining the wormhole throat \cite[Chapter 11]{Visser_Lorentzian_wormholes} in a well defined way. The proper radial distance $l$ in this spacetime is given by
\begin{equation}
    l(\rr) = \displaystyle\int_0^\rr \dfrac{d\rr'}{\Delta_\rr'(\rr)'^{1/2}} = \dfrac{1}{\sqrt{g^2-\A^2}}\tanh{^{-1}\left[\dfrac{\rr\sqrt{g^2-\A^2}}{\Delta_\rr(\rr)}\right]}
\end{equation}
Therefore after some manipulations, the coefficient of $d\Omega^2$ in the metric \eqref{C-metric} can be written as
\begin{equation}
    1 + r(l)^2 = 1 + \dfrac{1+g^2}{g^2-\A^2} \sinh{^2\left(l\sqrt{g^2-\A^2}\right)}
\end{equation}
for which, the minima of the function \(\sqrt{1+r(l)^2}\), and thus the wormhole throat is located at $l=0$ or $r=0$. Note that $r=0$ is a point in the spacetime, not a surface and hence two regions of the wormhole is marginally connected via a \emph{single point}. So it will not be a traversable wormhole even though there is no horizon for our choice of acceleration parameter.

\section{Geodesic Motion and Shadow Profile analysis}\label{section:geodesic}
\begin{figure}[H]
\centering
\subfloat{\includegraphics[width=.45\linewidth]{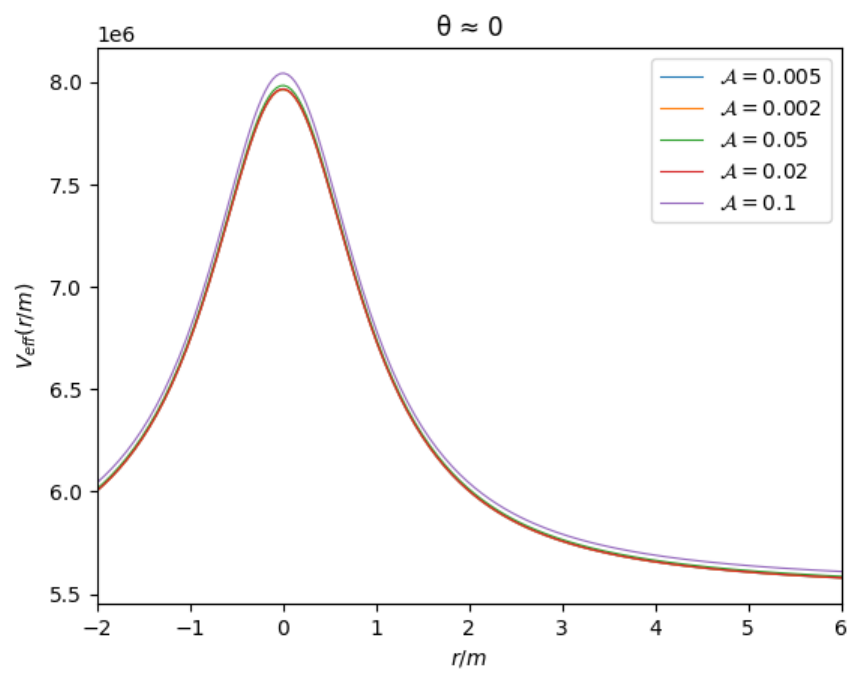}} \hspace{10mm}
\subfloat{\includegraphics[width=.45\linewidth]{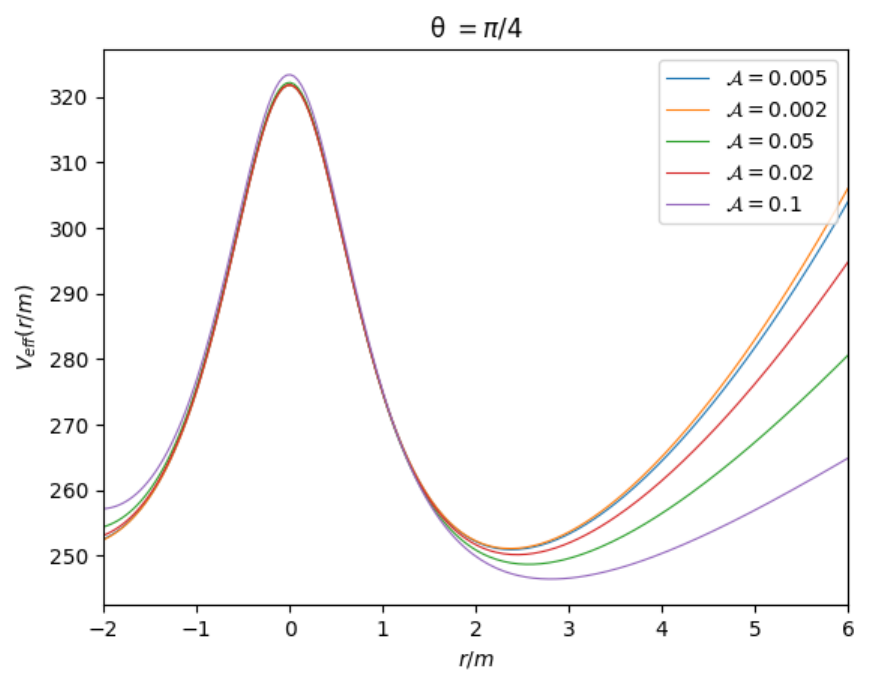}} \par
\subfloat{\includegraphics[width=.45\linewidth]{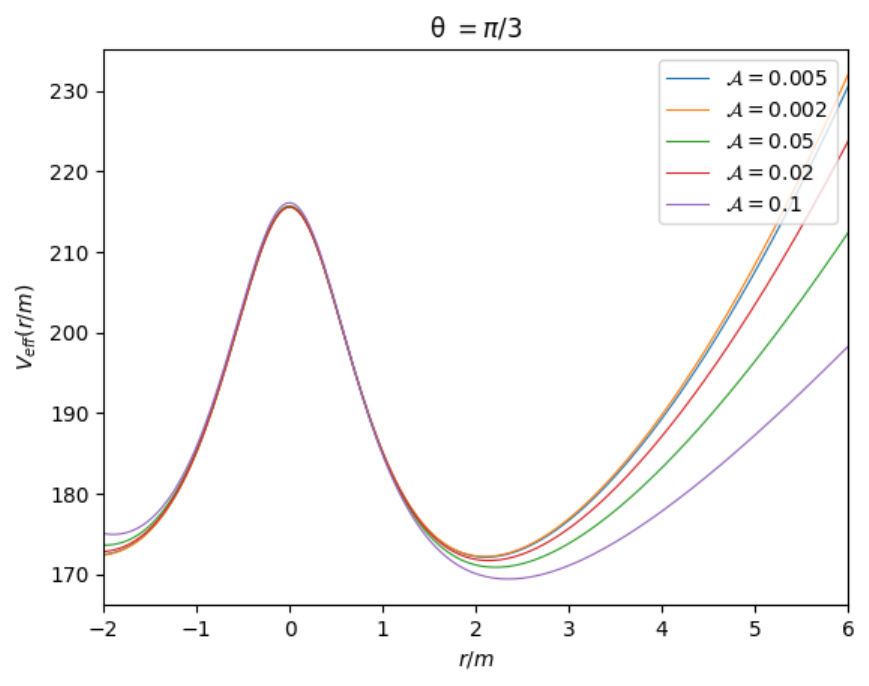}} \hspace{10mm}
\subfloat{\includegraphics[width=.45\linewidth]{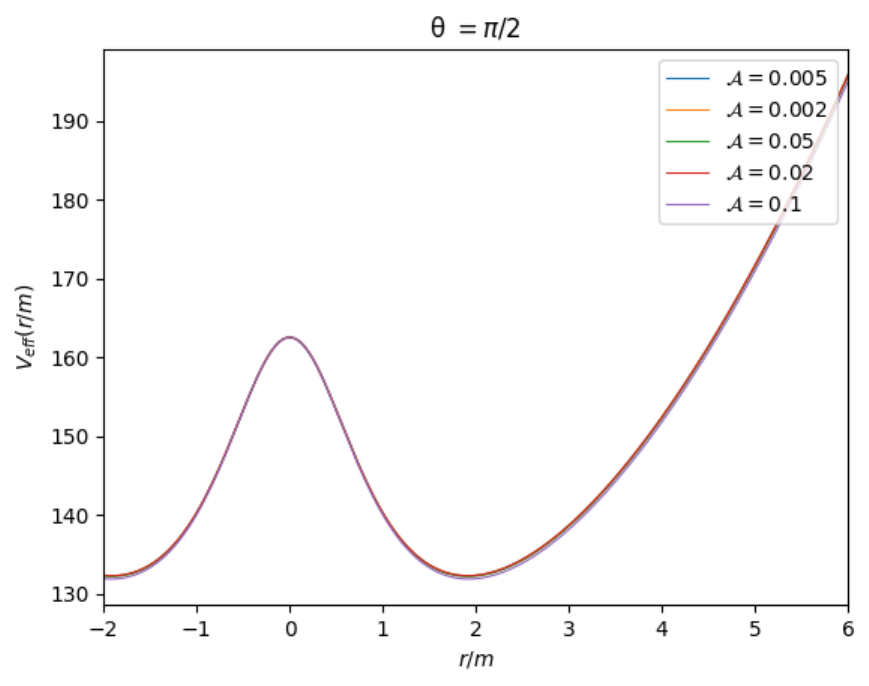}}\par
\subfloat{\includegraphics[width=.45\linewidth]{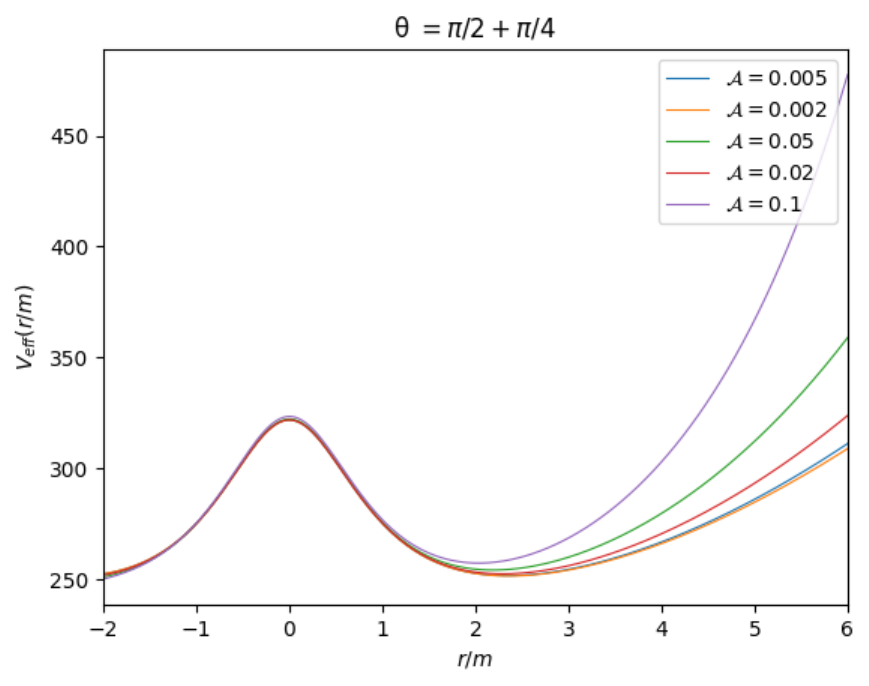}} \hspace{10mm}
\subfloat{\includegraphics[width=.45\linewidth]{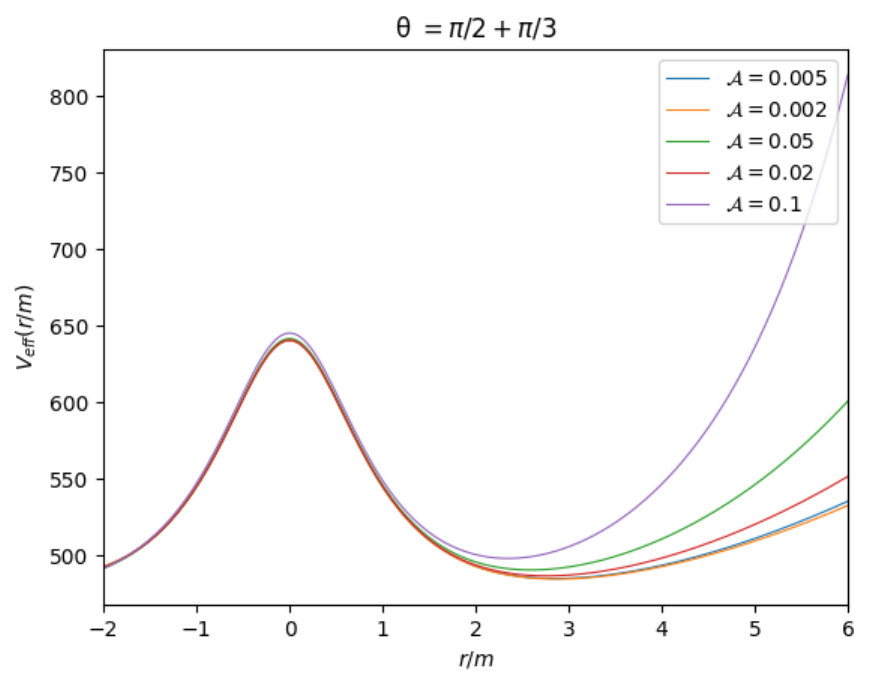}}
\caption{Variation of the effective potential with acceleration parameters for massless particles in different azimuthal angles $(\theta)$ for the following set of parameters: $g=1.5$, $\ell=7.0$.}
\label{fig:particle_eff_potential}
\end{figure}

\begin{figure}[H]
\centering
\subfloat{\includegraphics[width=.45\linewidth]{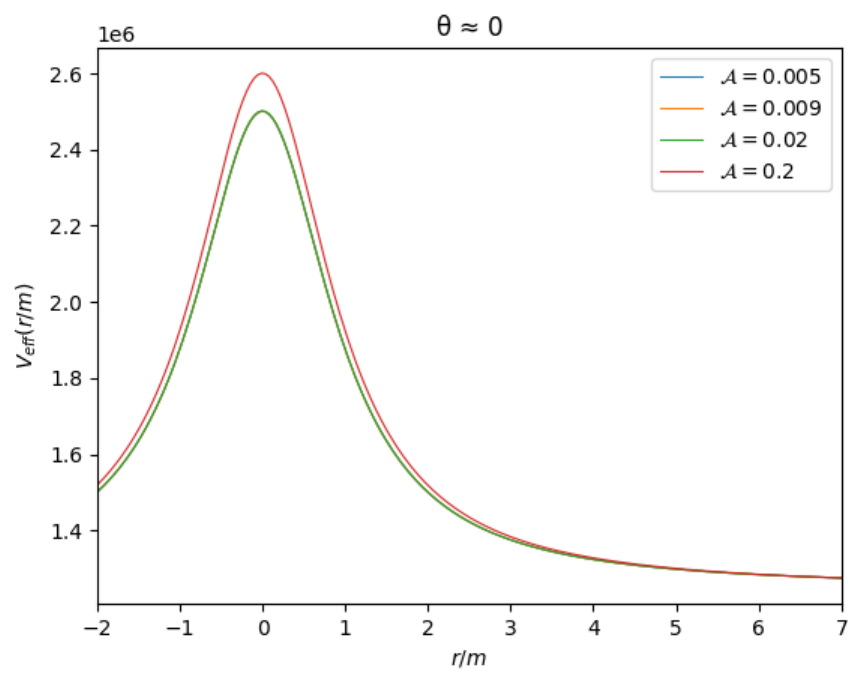}} \hspace{10mm}
\subfloat{\includegraphics[width=.45\linewidth]{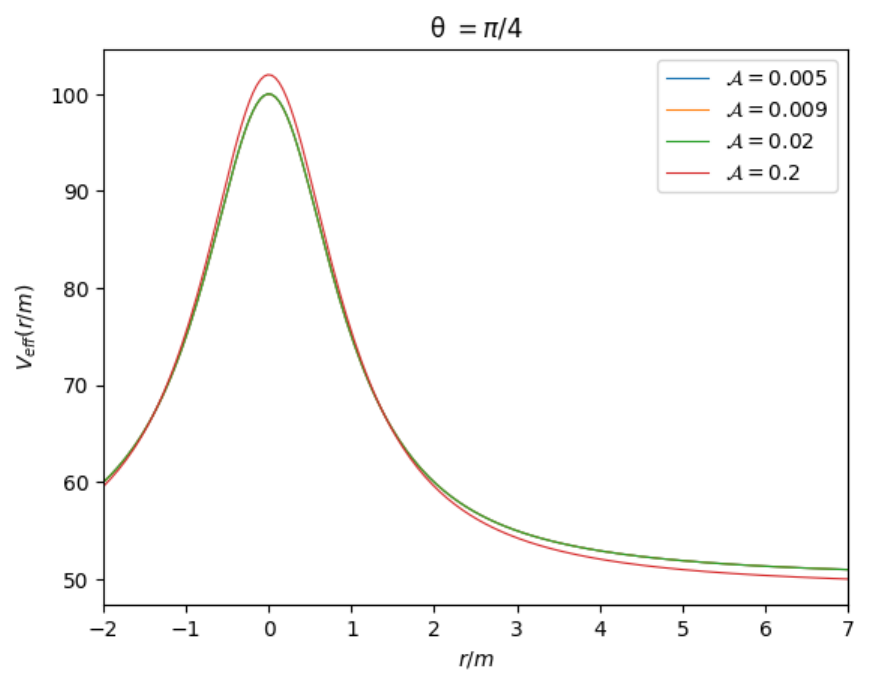}} \par
\subfloat{\includegraphics[width=.45\linewidth]{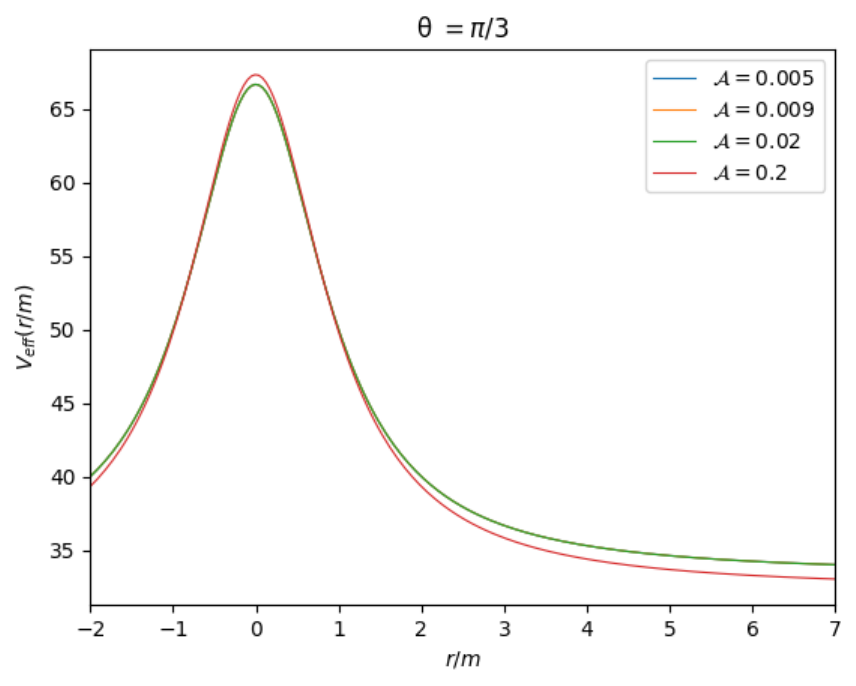}} \hspace{10mm}
\subfloat{\includegraphics[width=.45\linewidth]{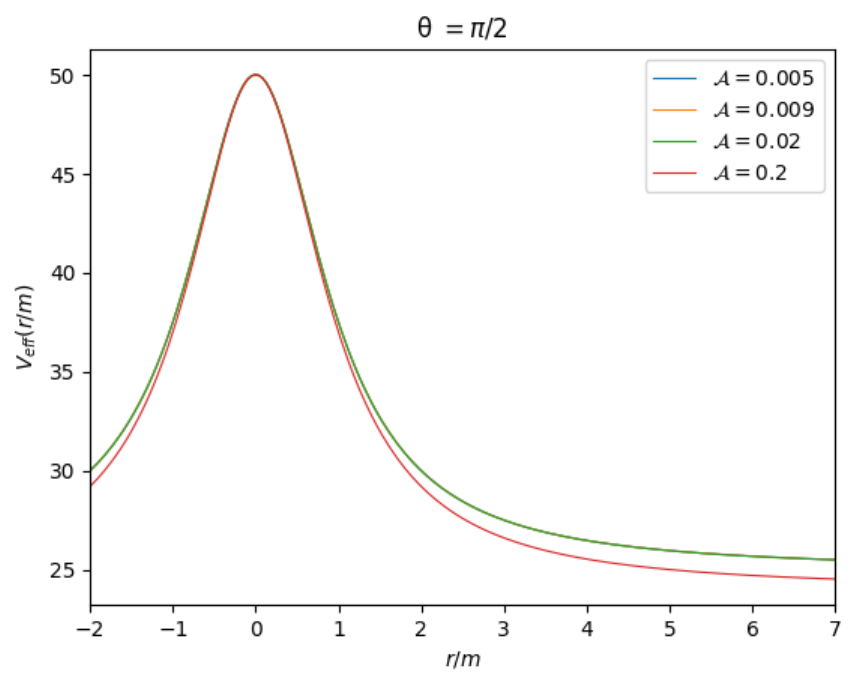}} \par
\subfloat{\includegraphics[width=.45\linewidth]{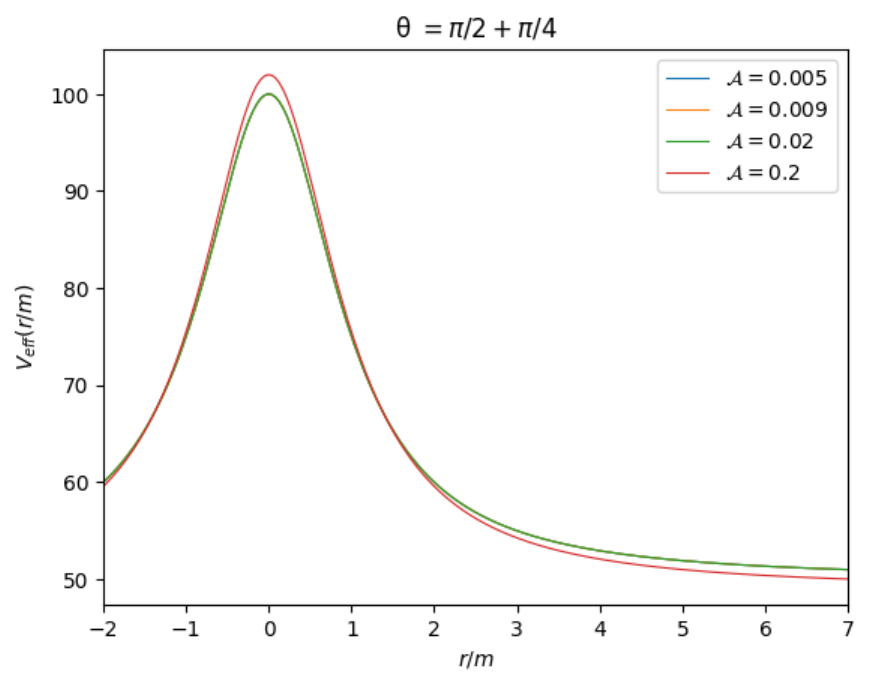}} \hspace{10mm}
\subfloat{\includegraphics[width=.45\linewidth]{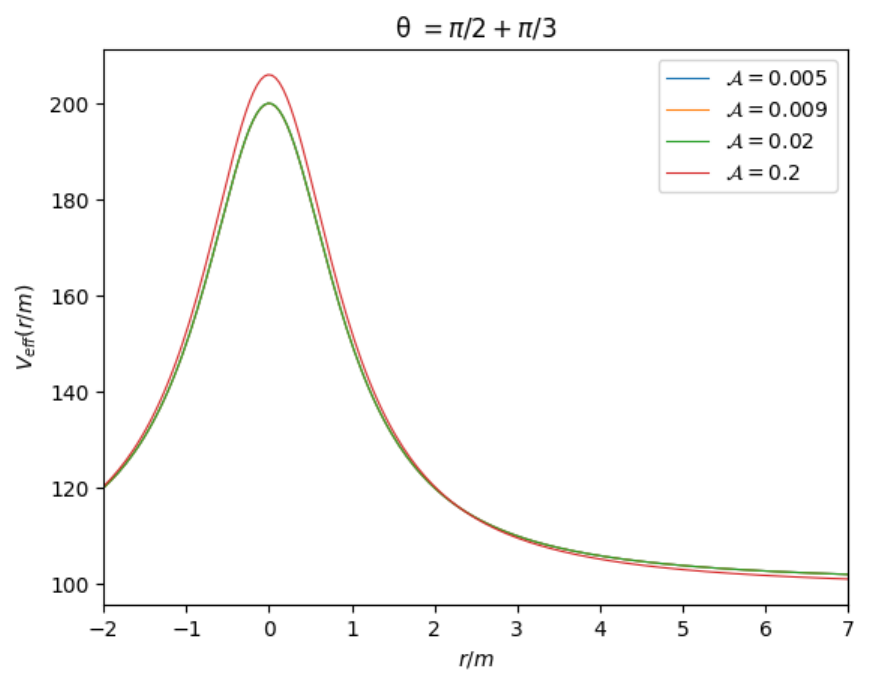}}
\caption{Variation of the effective potential with acceleration parameters for massless particles in different azimuthal angles $(\theta)$ for the following set of parameters: $g=1.0$, $\ell=5.0$.}
\label{fig:photon_eff_potential}
\end{figure}

The trajectory of a particle in this spacetime can be described by the curve $q^{\mu}(\lambda)=(\tau(\lambda),r(\lambda),x(\lambda),\varphi(\lambda))$, parameterized by $\lambda$ and the Lagrangian will be $\mathcal{L}=\frac{1}{2}g_{\mu \nu}\Dot{q}^{\mu}\Dot{q}^{\nu}$ (where overdot denotes the derivative with respect to the affine parameter). For the metric in Eq. \eqref{C-metric_solutions}, we can write
\begin{eqnarray}\label{Lagrangian}
    2\mathcal{L}=\dfrac{1}{(1+\A\rr x)^2}\left[(1+\A\rr^2x^2)\left(-\Delta_\rr(\rr)\Dot{\tau}^2+\dfrac{\Dot{\rr}^2}{\Delta_\rr(\rr)}\right)+(1+\rr^2)\left(\dfrac{\Dot{x}^2}{\Delta_x(x)}+\Delta_x(x)\Dot{\phi}^2\right)\right]= \epsilon
\end{eqnarray}
where $\epsilon=-1$ and $\epsilon=0$ correspond to timelike and lightlike geodesics respectively.
Then the conjugate momenta $p_{\mu}= \frac{\partial {\mathcal{L}}}{\partial \Dot{q}^{\mu}}$ will be 
\begin{eqnarray}
    p_t &=& -\dfrac{1+\A^2x^2}{(1+\A\rr x)^2}\Delta_\rr(\rr)\Dot{\tau} \equiv -E  \label{E}\\
    p_\rr &=& \dfrac{1+\A^2x^2}{(1+\A\rr x)^2}\dfrac{\Dot{\rr}}{\Delta_\rr(\rr)}\\
    p_x &=& \dfrac{1+\A^2x^2}{(1+\A\rr x)^2}\dfrac{\Dot{x}}{\Delta_x(x)}\\
    p_{\varphi} &=& \dfrac{1+\rr^2}{(1+\A\rr x)^2}-\Delta_x(x)\Dot{\varphi} \equiv \ell \label{ell}
\end{eqnarray}
Here the $\partial_t$ and $\partial_\varphi$ being the boost and rotational Killing vectors respectively, the conjugate momenta $p_t$ and $p_\varphi$ will be conserved and we define them as the energy $E$ and the angular momentum $\ell$ of the particle. The Euler-Lagrange equation $\dfrac{d}{d\lambda}\left(\dfrac{\partial \mathcal{L}}{\partial \Dot{q}^{\mu}}\right)-\dfrac{\partial \mathcal{L}}{\partial q^{\mu}}=0$ corresponding to the above Lagrangian \eqref{Lagrangian} will give the equations of motion. However, without directly deriving the geodesic equations, analyzing the effective potential can unveil certain properties regarding the behavior of a particle in this spacetime. Using the expressions of $\Dot{t}$ and $\Dot{\phi}$, we can rewrite eqn. \eqref{Lagrangian} as 
\begin{eqnarray}
    E^2-\dfrac{(1+\A^2x^2)^2}{(1+\A\rr x)^2}\Dot{\rr}^2-\dfrac{(1+\rr^2)(1+\A^2x^2)\Delta_\rr(\rr)}{(1+\A\rr x)^4\Delta_x(x)}\Dot{x}^2 = V_{eff}
\end{eqnarray}
where
\begin{eqnarray}
    V_{eff}(\rr) = \dfrac{\ell^2(1+\A^2x^2)\Delta_\rr(\rr)}{(1+\rr^2)\Delta_x(x)}-\dfrac{\epsilon (1+\A^2 x^2)\Delta_\rr(\rr)}{(1+\A\rr x)^2}
\end{eqnarray}
We will analyse the near-throat behaviour of the effective potential $(V_{eff})$ as the astrophysically interesting phenomena takes place near the throat of the wormhole. For massive particles, the effective potential reaches an asymptotic value 
$(g^2-\A^2)(1+\A^2x^2)\left[\frac{\ell^2}{1-x^2}+\frac{1}{\A^2x^2}\right]$, or in other words it will reach a constant value at large $r$ for timelike geodesic except when $\theta = 0$ and $\theta = \pi/2$. In Fig. \ref{fig:particle_eff_potential}, the plots depict the variation of the effective potentials, a massive particle will experience, for different acceleration parameters $\A$. For $\theta=0$ and $\theta=\pi/2$, there is an unstable equilibrium point at $r = 0$, but as for this spacetime, $r=0$ is merely a point, there will be no closed timelike geodesic at north pole and at equatorial plane. In between these two regions, there can exist an unstable equilibrium for massive particles that reaches close to $r=0$ at slow acceleration limit. All of these analysis is also in agreement with the expression of the first derivative of effective potential at $r=0$ for $\epsilon=-1$, i.e., $2\A x(1-x^2-g^2(x^2-1))$. Furthermore, a minimum exists in the effective potential for all values of $\A$ except when $\theta \approx 0$. Therefore, for massive particles, a bound state is present everywhere except near the north pole.
For photons $\epsilon=0$, $V_{eff}(\rr)$ possess only one maxima at throat, $\rr=0$, independent of $x$ as evident from Fig. \ref{fig:photon_eff_potential}. So, there is no stable or unstable orbit for the photon in contrast to the timelike case. In both of the cases, the maximum value of the effective potential increase from equatorial plane to north pole gradually. It is noteworthy that for both of the cases, in small acceleration limit $(\A \rightarrow 0)$, the behaviour of effective potential is nearly same for all values of $\A$ for a particular azimuth.

Now, using the Hamiltonian of the particle, obtained by the Legendre transformation of the Lagrangian $\mathcal{L}$, we can also write the Hamilton-Jacobi equation $\mathcal{H} \left(q,\frac{\partial \mathcal{S}}{\partial q}\right)+ \frac{\partial \mathcal{S}}{\partial \lambda}=0$ and substituting the following ansatz of Hamilton's principle function for a photon, $\mathcal{S}=-E\tau+\mathcal{S}_\rr(\rr)+\mathcal{S}_x(x)+\ell \varphi$, the equation separates as 
\begin{eqnarray}\label{HJ_1_2}
    \dfrac{\partial \mathcal{S}_x}{\partial x} &=& \pm \left[\dfrac{\kappa}{\Delta_x(x)(1+\A^2x^2)}-\dfrac{\ell^2}{\Delta^2_x(x)}\right]^{1/2}\\
    \dfrac{\partial \mathcal{S}_\rr}{\partial \rr} &=& \pm \left[\dfrac{E^2}{\Delta^2_\rr(\rr)}-\dfrac{\kappa}{\Delta_\rr(\rr)(1+\rr^2)}\right]^{1/2}
\end{eqnarray}
and this yields a Carter--like separation constant $\kappa$. Now using $p_{\mu}= \frac{\partial \mathcal{S}}{\partial q^{\mu}}$ along with the expressions of conjugate momenta from Lagrangian, we obtain
\begin{eqnarray}
    \Dot{\tau} &=& \dfrac{E(1+\A\rr x)^2}{(1+\A^2x^2)\Delta_\rr(\rr)} \label{tau_dot}\\
    \Dot{\rr} &=& \pm \dfrac{(1+\A\rr x)^2}{(1+\A^2x^2)}\left[E^2-\dfrac{\kappa \Delta_\rr(\rr)}{1+\rr^2}\right]^{1/2} \label{r_dot}\\
    \Dot{x} &=& \pm  \dfrac{(1+\A\rr x)^2}{(1+\rr^2)}\left[\dfrac{\kappa \Delta_x(x)}{1+\A^2x^2}-\ell^2\right]^{1/2} \label{x_dot}\\
    \Dot{\varphi} &=& \dfrac{\ell(1+\A\rr x)^2}{(1+\rr^2)\Delta_x(x)} \label{phi_dot}
\end{eqnarray}
The existence of the geodesics require the positiveness of the terms under the square roots in Eqs. \eqref{r_dot} and \eqref{x_dot} which in turn imply $\kappa \geq 0$ and $\kappa > \ell^2$. Now these conditions indicate that $\Dot{\rr}$ and $\Ddot{\rr}$ cannot vanish simultaneously and same goes for $\Dot{x}$ and $\Ddot{x}$ also. So there neither exists aspherical photon surface nor any photon cone in this spacetime. Therefore \emph{no} shadow will be cast by the spacetime due to the non-existence of any special photon surface. We can also understand this as the effect of the absence of any photon orbit in this spacetime.

From the analysis presented in this section, two key points emerge: Firstly, for slow acceleration, the nature of the effective potential remains ``almost'' indistinguishable for different values of acceleration for both massive and massless particles near the astrophysically most enriched region of the wormhole, i.e., near the throat. Secondly, the spacetime lacks the ability to cast shadow, making it challenging to observationally discern this class of spacetime by studying particle behavior. Hence we will now turn ourselves to the study of (normal and) quasinormal modes of this spacetime, which may lead to phenomenologically interesting possibilities.

\section{Scalar waves}\label{sec:ScalarPerturbation}

Let us consider a free massless scalar field non--minimally coupled to this geometry, while also not interacting with the background scalar field $\Phi$. Additionally, assume that this field does not contribute to the stress-energy tensor of the background system. Thus, it will merely act as a perturbation to the system, allowing us to analyze the effect of the background geometry on this free scalar field. The conformally invariant Klein--Gordon equation for this field is given by 
\begin{equation}
    \dfrac{1}{\sqrt{-\Bar{g}}}\partial_{\mu}\left(\sqrt{-\Bar{g}}\Bar{g}^{\mu\nu}\partial_{\nu}\Bar{\phi}\right) + \alpha\Bar{R} \Bar{\phi} = 0 \label{KleinGordon_equation}
\end{equation}
where $\alpha$ is generally viewed either as a coupling factor of $\Bar{\phi}\Bar{R}$ type coupling or it takes value $-\frac{1}{6}$ when the metric is used without its conformal factor \cite{charged_c_metric_QNM&superradiance}. In that case, $\Bar{g}^{\mu\nu}$ and $\Bar{R}$ are the conformal metric and its Ricci scalar respectively. Now at this point, this is worth emphasizing that the line element \eqref{C-metric} doesn't admit any rank-$2$ conformal Killing--Yano tensor, which makes it impossible to perform separation of variable of the Klein--Gordon equation \eqref{KleinGordon_equation} leading to enormous difficulties in solving it. In this situation, the entire line element \eqref{C-metric} can be redefined by scaling through the conformal factors of the $2$D line element \(\left(-\Delta_\rr(\rr)d\tau^2+\Delta_\rr(\rr)^{-1}dr^2\right)\) as
\begin{equation}
    d\Tilde{s}^2 = -\left(1+g^2+(g^2-\A^2)\rr^2\right) d\tau^2 + \dfrac{d\rr^2}{1+g^2+(g^2-\A^2)\rr^2} + \dfrac{1+\rr^2}{1+\A^2x^2} \left(\dfrac{dx^2}{1-x^2} + (1-x^2) d\varphi^2\right) \label{rescaled_C-metric1}
\end{equation}
Now we perform another coordinate transformation, $t\to\frac{\zeta}{\sqrt{1+g^2}}$ and $\rr\to\sqrt{1+g^2}\rho$ so that the metric takes the form
\begin{equation}
    d\Tilde{s}^2 = -\left(1+\varepsilon\rho^2\right) d\zeta^2 + \dfrac{d\rho^2}{1+\varepsilon\rho^2} + \dfrac{1+a^2\rho^2}{1+\A^2x^2} \left(\dfrac{dx^2}{1-x^2} + (1-x^2) d\varphi^2\right) \label{rescaled_C-metric2}
\end{equation}
where $\varepsilon=g^2-\A^2$ and $a^2=1+g^2$. This metric will now admit a Killing--Yano $2$-form \(\frac{(1+a^2\rho^2)^{3/2}}{1+\A^2x^2} dx \wedge d\varphi\), along with five other rank-$2$ Killing tensors, which make it possible to perform the following decomposition of the scalar field,
\begin{equation}
    \Bar{\phi}(\zeta,\rho,x,\varphi) = \dfrac{(1+a\A \rho x)^2}{1+\A^2x^2} \phi(\zeta,\rho,x,\varphi) \sim \dfrac{(1+a\A \rho x)^2}{1+\A^2x^2} \displaystyle\int d\omega ~ e^{-i\omega\zeta+iM\varphi} \dfrac{\psi_{LM}(\rho)\Theta_{LM}(x)}{\sqrt{1+\varepsilon\rho^2}} \label{phi_decomposition}
\end{equation}
where $L$ denotes the excitation level of the scalar field and $M$ is a quantum number representing the direction of the orientation of that field in spacetime. From now on, the subscript $LM$ will be dropped for the sake of brevity.

The ansatz \eqref{phi_decomposition} will now lead to the separation of the Eq. \eqref{KleinGordon_equation} into a radial part and an angular part. In the limit of extremely low acceleration $(\A\rightarrow 0)$, the angular component simplifies to the well-known associated Legendre polynomial $Y_{LM}(x)$ due to the system's spherical symmetry. We will now analyse these two separated equations below by confining ourselves mostly in the low acceleration limit mainly for two reasons:
\begin{enumerate}
    \item The experimental bound on the acceleration parameter of an accelerating Kerr--Newman black hole, from the shadow of M87* \cite{bound1}, suggests that celestial bodies either have extremely small acceleration or zero acceleration.
    \item For the original spacetime \eqref{C-metric}, as mentioned earlier, the position of the spatial infinities inversely varies with the acceleration parameter. Thus it is convenient and physically intuitive to take small acceleration limit, so that the spatial infinities gets mapped to the \textit{mathematical infinities} $\pm\infty$.
\end{enumerate}

\subsection{Angular part}
The decomposition \eqref{phi_decomposition} along with Eqs. \eqref{KleinGordon_equation} and \eqref{rescaled_C-metric2} will give the angular equation as follows.
\begin{equation}
    \left(1-x^2\right) \dfrac{d^2\Theta}{dx^2} - 2x \dfrac{d\Theta}{dx} + \left[\dfrac{\xi}{\left(1+\A^2x^2\right)} - \dfrac{M^2}{1-x^2} - \dfrac{\left(1+\A^2\right)\left(1-\A^2x^2\right)}{3\left(1+\A^2x^2\right)^2}\right] \Theta = 0 \label{angular_equation}
\end{equation}
where $6\xi$ is the separation constant of angular and radial part equation. This equation is almost like the equation of an associated Legendre polynomial with eigenvalue $6\xi$, except for the fact that there is an extra parameter for acceleration here. At this point, one might ask the question how to solve this equation, as there is no standard technique to find the exact analytical solution of this. Similar (but not exactly same) angular equation arises while doing the QNM analysis of ordinary C--metrics. It is popular in the literature \cite{qnm1,qnm2,qnm3,qnm4,qnm5,qnm6,qnm7,qnm8,qnm9,qnm10} to find the numerical solution of that equation by imposing the appropriate boundary conditions: $\Theta(x) \sim e^{\pm Mx}$ as $x\to\pm1$. But in our case, pseudospectral method \cite{QNMspectral_paper} along with these boundary conditions producing poor and sometimes, null results. Therefore let us turn our attention to an approximate analytical result, which is not only physically more intuitive, but also give accurate results in proper physical limit(s).

In the low acceleration limit, a perturbative series expansion of the solution of Eq. \eqref{angular_equation} can be performed and the eigenvalue $\xi$ for a given $M$ and $\A$ can be calculated upto $2$nd order. We will mostly confine ourselves to the expression of $\xi$ only, rather than the solution itself; because, the solution $\Theta(x)$ can be affected by the source terms, but the eigenvalue can't and thus it carries the inherent properties of the system. Besides these, $\xi$ is the only term that will be needed to calculate the final (normal and) quasinormal modes for the system. The detailed perturbative analysis of the equation has been given in Appendix \ref{perturbative_series_angular}.

Second order expansion of $\xi$ reads as
\begin{equation}
    \xi = L(L+1) + \dfrac{1}{3} + \A^2 \left[2 + \dfrac{(2L+1)(L-M)!}{2(L+M)!} \left(L(L+1) -\dfrac{2}{3}\right) \displaystyle\int_{-1}^{+1} \chi'^2 P_{LM}(\chi')^2 d\chi'\right] + \mathcal{O}(\A^4) \label{xi_expression}
\end{equation}
As the rescaled line element \eqref{rescaled_C-metric1} and \eqref{rescaled_C-metric2} only contains $\A^2$, the even power of $\A$, so Eq. \eqref{xi_expression} also contains only the even powers of $\A$. Now the definite integral in Eq. \eqref{xi_expression} can be evaluated in terms of $L$ and $M$ in general using Gaunt's formula \cite{GauntFormula1,GauntFormula2,GauntFormula3}, but that will make the expression unnecessarily messy. Rather it is easier to compute the integral for $M=0$. For $L\ge2$, the integral reduces to
\begin{equation}
    \displaystyle\int_{-1}^{+1} \chi'^2 P_{LM}(\chi')^2 d\chi' = \dfrac{2}{3(2L+1)} \left[1 + \dfrac{2L(L+1)}{(2L+3)}\right]
\end{equation}
and using this, in eikonal limit, $\xi$ for $M=0$ can be written as
\begin{equation}
    \xi \simeq L^2 \left[1+\dfrac{\A^2L}{9}\right] + \mathcal{O}(\A^4)
\end{equation}

\subsection{Radial part}
\begin{figure}[H]
\centering
\subfloat{\includegraphics[width=\linewidth]{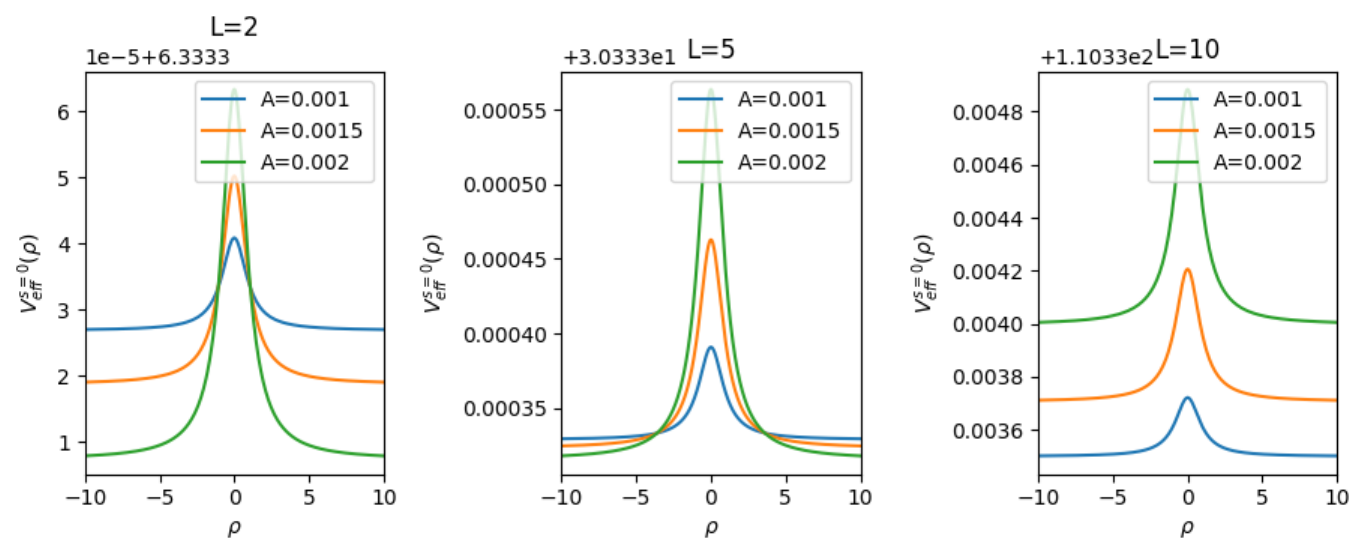}}
\caption{$V_{eff}^{s=0}(\rho)$ for different acceleration parameters and for a fixed $g=1.0$ with $M=0$.}
\label{fig:scalar_eff_potential}
\end{figure}

Substitution of the ansatz \eqref{phi_decomposition} in Eq. \eqref{KleinGordon_equation} decouples the radial equation as
\begin{equation}
     \dfrac{d^2\psi}{d\rho_*^2} + \left(\varpi^2-\dfrac{1}{\varepsilon}V_{eff}^{s=0}\right)\psi = 0 \label{master_equation}
\end{equation}
where $\varpi$ is the reduced frequency, defined by \(\varpi=\omega/\sqrt{\varepsilon}\) and $\rho_*$ is the tortoise coordinate, defined by
\begin{equation}
    \rho_* = \sqrt{\varepsilon} \displaystyle\int_0^\rho \dfrac{d\rho'}{1+\varepsilon\rho'^2} = \tan{^{-1}\left(\sqrt{\varepsilon}\rho\right)} \label{tortoise_rho*}
\end{equation}
and $V_{eff}^{s=0}(\rho)$ is the scalar (corresponds to $s=0$) effective potential, given by
\begin{equation}
    V_{eff}^{s=0}(\rho) = \dfrac{1+\varepsilon\rho^2}{3\left(1+a^2\rho^2\right)^2} \left[a^2\rho^2(\varepsilon+3\xi)+3\xi+a^2-\varepsilon-a^4\rho^2\right] \label{scalar_effective_potential}
\end{equation}
where the expression for $\xi$ is given in Eq. \eqref{xi_expression}. This potential has a global maxima at the throat of the wormhole. Besides this, for $\xi < \frac{1}{3}a^2-\varepsilon = \A^2-\frac{2}{3}g^2+\frac{1}{3}$, it will also have one minima on each side of the wormhole. But as the acceleration (and possibly $g$ also) is considered to be small, resulting $\A^2-\frac{2}{3}g^2+\frac{1}{3}$ to be small (or maybe even negative), so in \emph{eikonal limit}, the potential will have only one stationary point at $\rho=0$ and that is maxima. Also this is worth noting that the potential \eqref{scalar_effective_potential} reaches an asymptotic value of $\frac{\varepsilon(-a^2+\varepsilon+3\xi)}{3a^2}$ as $\rho\to\infty$, which is an artifact of the masslessness of perturbing scalar field. The variation of the scalar effective potential over the radial distance for different choices of acceleration and for a fixed $g$ has been shown in Fig. \ref{fig:scalar_eff_potential}. 

The boundary conditions for the master equation \eqref{master_equation} can be chosen conventionally: the waves are outgoing at the spatial infinities on both sides of the wormhole or no waves come from either the left or the right infinity. However, since the tortoise coordinate doesn't extend from $-\infty$ to $+\infty$, the WKB approximation, even with Padé approximations \cite{WKB_paper}, cannot be applied. Alternatively, the use of \emph{Mathematica} package \texttt{QNMspectral} \cite{QNMspectral_paper} produces extremely poor and insignificant results. Therefore it is better to seek for an approximated analytic result to have some physical insight of the system.

Let us consider the eikonal limit (large $\xi$) of the system, such that $\xi>>a^2$, which immediately follows the fact that $\varepsilon+3\xi$ and $3\xi+a^2-\varepsilon$ can safely be approximated as $3\xi$. With solely this minimal amount of realistic consideration, the scalar effective potential can be written as
\begin{eqnarray}
    \dfrac{1}{\varepsilon} V_{eff}^{s=0}(\rho) &\simeq& \dfrac{\xi\left(1+\varepsilon\rho^2\right)}{\varepsilon\left(1+a^2\rho^2\right)} \label{Veff_intermediate1} \\
    &=& \dfrac{\xi}{a^2} + \left(\dfrac{1}{\varepsilon}-\dfrac{1}{a^2}\right) \dfrac{\xi}{1+a^2\rho^2}
\end{eqnarray}
Now using Eq. \eqref{tortoise_rho*} and the transformation rules, the effective potential reduces after some algebraic and trigonometric manipulations as follows:
\begin{equation}
    \dfrac{1}{\varepsilon} V_{eff}^{s=0}(\rho_*) \simeq \dfrac{\xi}{\left(1+g^2\right)-\frac{1}{2}\left(1+\A^2\right)\left(1+\cos{2\rho_*}\right)} \label{scalar_effective_potential_rho*}
\end{equation}
Using this potential, the solution of the master equation \eqref{master_equation} takes the form
\begin{eqnarray}
    \psi(\rho_*) = &c_1& \Hl\left[\dfrac{1+g^2}{1+\A^2},\dfrac{\xi-\left(1+g^2\right)\varpi^2}{4\left(1+\A^2\right)};-\dfrac{\varpi}{2},\dfrac{\varpi}{2},\dfrac{1}{2},\dfrac{1}{2};\cos{^2\rho_*}\right] + \nonumber \\
    &c_2& \Hl\left[\dfrac{1+g^2}{1+\A^2},\dfrac{\xi-\left(1+g^2\right)\left(\varpi^2-1\right)}{4\left(1+\A^2\right)};\frac{1-\varpi}{2},\dfrac{1+\varpi}{2},\dfrac{3}{2},\dfrac{1}{2};\cos{^2\rho_*}\right] \cos{\rho_*} \label{master_soltuion_Heun}
\end{eqnarray}
where $\Hl$ is the \textit{local Heun function} (see Appendix \ref{Heun_function}), which is regular at the spatial infinity $(\rho_*\to\pm\pi/2)$, but not at the throat $(\rho_*=0)$ of the wormhole. It is intriguing to note that although the scalar effective potentials \eqref{scalar_effective_potential} and \eqref{scalar_effective_potential_rho*} exhibit symmetry about the throat, the system only permits even solutions, ruling out odd solutions regardless of the value of $\varpi$. Additionally, observing that $\psi(\rho_*)=c_1$ at spatial infinities is a consequence of the normalization of the Heun function, as detailed in Appendix \ref{Heun_function}. It is worth emphasizing that the solution \eqref{master_soltuion_Heun} is sufficiently generalized, and we have not employed any approximation other than the eikonal limit, which is both justified and self-explanatory. Now to ensure non-zero finiteness of the solutions at the throat, it necessitates the regularization of the local Heun function at that point.

For this purpose, we use Sleeman--Meixner--Sch{\"a}fke theorem \cite{Sleeman_Heun@z1,meixner2013mathieusche}. It states that for the local Heun function \(\Hl(\Tilde{a},\Tilde{q};\Tilde{\alpha},\Tilde{\beta},\Tilde{\gamma},\Tilde{\delta};\Tilde{z})\) to be regular at $\Tilde{z}=1$, it requires
\begin{equation}
    \mu_0 + \K_{j=1}^\infty \dfrac{\nu_j \lambda_j}{\mu_j} = 0 \label{Sleeman--Meixner--Schafke_theorem}
\end{equation}
where $\operatornamewithlimits{K}$ is the notation for standard continued fraction and the rest of the symbols explained below:
\begin{eqnarray}
    \mu_j &=& -\left[j^2(1+\Tilde{a}) + j\{\Tilde{a}(\Tilde{\gamma}+\Tilde{\delta}-1)+(\Tilde{\alpha}+\Tilde{\beta}-\Tilde{\delta})\} + \Tilde{q}\right] , \\
    \nu_j &=& \Tilde{a}j(j+\Tilde{\gamma}-1) , \\
    \lambda_j &=& -(j+\Tilde{\alpha}-1)(j+\Tilde{\beta}-1)
\end{eqnarray}
We have used this theorem \eqref{Sleeman--Meixner--Schafke_theorem} for each of the local Heun functions in Eq. \eqref{master_soltuion_Heun} by truncating the continued fraction upto a value (say $N$) and found the root for $\varpi$. Truncation of the continued fraction makes it a rational polynomial and the number of roots are obviously directly proportional to $N$. The value of $N$ can be chosen such that after which the rate of change of desired number of roots go below the tolerance one wants (we chose $N=15$ and got sufficiently accurate roots). Thus one finds two sets of quantized $\varpi$ corresponding to each of the solutions in Eq. \eqref{master_soltuion_Heun}.

Requiring $\psi(\rho_*)$ to be finite at the throat necessitates both solutions of Eq. \eqref{master_equation} to be regular at $\rho_*=0$. Therefore only the common values of $\varpi$ from the aforementioned two sets are permissible. Hence under the previously outlined boundary conditions, no common modes exist in the system, leading to divergence of the field at the throat and disconnection between the field in the two halves of the wormhole. Also $\varpi$ remains unquantized, indicating that these boundary conditions yield \textit{continuous spectra}.

\begin{itemize}
    \item \textbf{Possibilities with a different boundary condition:} As the spacetime doesn't have asymptotic flatness and we also don't get interesting result from perturbation theory, there is an avenue to explore other possibilities of the boundary conditions; such as, the field derivative may decay so much, such that it vanishes at the spatial infinity, which is also supported from the asymptotic behaviour of the scalar effective potential.
\end{itemize}

In that case, the integration coefficient $c_2$ in Eq. \eqref{master_soltuion_Heun} must vanish \cite{Heun_derivative}. Then the resultant first few (normal and quasinormal) frequencies for different mode values (with $M=0$) obtained from the Sleeman--Meixner--Sch{\"a}fke theorem are tabulated in Table \ref{field_derivative_zero_table1}-\ref{field_derivative_zero_table4}. (The potential \eqref{Veff_intermediate1} doesn't seem to result in growing modes making the system unstable. So from $\pm\omega_R \pm i\omega_I$ and $\pm\omega_R \mp i\omega_I$ type of modes, we have only chosen the positive real and negative imaginary ones for the sake of conveniences).
As evident from the following tables, the real part of the frequency consistently surpasses the imaginary part, regardless of the value of $\mathcal{A}$. Moreover, both components of the frequency show an increasing trend as the acceleration parameter rises. The increase in imaginary part indicates that the system is damping due to the leakage of waves at two extreme ends, though we have not explicitly considered the outgoing nature of waves at the two spatial infinities in our boundary conditions. This implies that our current boundary conditions in turn imply the outgoing nature of some part of the waves at excited states (i.e. $n>0$) at the spatial infinities. With the increase in overtone number, $\Re{(\omega)}$ increases much faster than $\Re{(\omega)}$, that means that the field holds the excitation more and more when it goes to the higher excited states. 
\begin{table}[!htbp]
\centering
\caption{For $\mathcal{A} = 0.1$ and $g = 1.0$}
\begin{tabular}{|c|c|c|c|c|c|}
\hline
n & $L=400$ & $L=500$ & $L=700$ & $L=850$ & $L=1000$ \\
\hline
$0$ & $(22.7072,0.0000)$ & $(22.7147,0.0000)$ & $(22.7202,0.0000)$ & $(22.7219,0.0000)$ & $(22.7228,0.0000)$ \\
\hline
$1$ & $(25.3524,-10.7430)$ & $(28.1353,-16.2397)$ & $(36.3661,-28.1010)$ & $(44.0305,-37.4159)$ & $(52.5687,-47.0839)$ \\
\hline
$2$ & $(28.6404,-17.4866)$ & $(33.9600,-25.1759)$ & $(47.0434,-40.9291)$ & $(58.3024,-53.2943)$ & $(70.5162,-66.2251)$ \\
\hline
$3$ & $(31.4040,-24.5303)$ & $(39.0125,-33.4793)$ & $(56.2685,-52.1527)$ & $(70.6284,-67.0301)$ & $(86.0250,-82.7051)$ \\
\hline
$4$ & $(37.5818,-33.7568)$ & $(47.9323,-44.5644)$ & $(70.7409,-67.6400)$ & $(89.4471,-86.2645)$ & $(109.3890,-105.9930)$ \\
\hline
$5$ & $(50.8534,-47.6931)$ & $(65.6663,-62.1306)$ & $(97.9158,-93.3085)$ & $(124.2040,-118.6200)$ & $(152.1610,-145.4940)$ \\
\hline
$6$ & $(84.1677,-75.0804)$ & $(109.1120,-97.4371)$ & $(163.2190,-145.8840)$ & $(207.2450,-185.2830)$ & $(254.0300,-227.1440)$ \\
\hline
\end{tabular}
\label{field_derivative_zero_table1}
\end{table}

\begin{table}[!htbp]
\centering
\caption{For $\mathcal{A} = 0.02$ and $g = 1.0$}
\begin{tabular}{|c|c|c|c|c|c|}
\hline
n & $L=400$ & $L=500$ & $L=700$ & $L=850$ & $L=1000$ \\
\hline
$0$ & $(22.8067,0.0000)$ & $(22.8190,0.0000)$ & $(22.8269,0.0000)$ & $(22.8294,0.0000)$ & $(22.8308,0.0000)$ \\
\hline
$1$ & $(24.4459,-7.9773)$ & $(25.8862,-11.7363)$ & $(30.3779,-19.7167)$ & $(34.5869,-25.6826)$ & $(39.1919,-31.5608)$ \\
\hline
$2$ & $(26.3106,-13.1878)$ & $(29.6462,-18.9416)$ & $(37.6806,-29.8498)$ & $(44.3097,-37.7495)$ & $(51.2140,-45.5290)$ \\
\hline
$3$ & $(27.6234,-19.6083)$ & $(32.8536,-26.2352)$ & $(44.0091,-38.9942)$ & $(52.7147,-48.3691)$ & $(61.5973,-57.6848)$ \\
\hline
$4$ & $(32.2453,-27.9533)$ & $(39.5535,-35.8175)$ & $(54.5812,-51.3411)$ & $(66.0692,-62.9423)$ & $(77.6873,-74.5612)$ \\
\hline
$5$ & $(43.0946,-40.0503)$ & $(53.6761,-50.4451)$ & $(75.0940,-71.2623)$ & $(91.3257,-86.9473)$ & $(107.6810,-102.7120)$ \\
\hline
$6$ & $(71.0397,-63.2988)$ & $(88.9217,-79.3415)$ & $(124.9430,-111.6150)$ & $(152.1710,-135.9930)$ & $(179.5740,-160.5200)$ \\
\hline
\end{tabular}
\label{field_derivative_zero_table2}
\end{table}

\begin{table}[!htbp]
\centering
\caption{For $\mathcal{A} = 0.005$ and $g = 1.0$}
\begin{tabular}{|c|c|c|c|c|c|}
\hline
n & $L=400$ & $L=500$ & $L=700$ & $L=850$ & $L=1000$ \\
\hline
$0$ & $(22.8103,0.0000)$ & $(22.8229,0.0000)$ & $(22.8309,0.0000)$ & $(22.8335,0.0000)$ & $(22.8350,0.0000)$ \\
\hline
$1$ & $(24.4158,-7.8635)$ & $(25.8022,-11.5408)$ & $(30.1272,-19.3268)$ & $(34.1697,-25.1195)$ & $(38.5766,-30.7976)$ \\
\hline
$2$ & $(26.2188,-12.9970)$ & $(29.4679,-18.6599)$ & $(37.2667,-29.3315)$ & $(43.6687,-37.0059)$ & $(50.3030,-44.5182)$ \\
\hline
$3$ & $(27.4642,-19.3914)$ & $(32.5874,-25.9097)$ & $(43.4550,-38.3835)$ & $(51.8816,-47.4833)$ & $(60.4316,-56.4711)$ \\
\hline
$4$ & $(32.0166,-27.6997)$ & $(39.1862,-35.4283)$ & $(53.8443,-50.5905)$ & $(64.9745,-61.8420)$ & $(76.1663,-73.0441)$ \\
\hline
$5$ & $(42.7595,-39.7186)$ & $(53.1474,-49.9279)$ & $(74.0493,-70.2506)$ & $(89.7817,-85.4571)$ & $(105.5420,-100.6520)$ \\
\hline
$6$ & $(70.4714,-62.7885)$ & $(88.0300,-78.5419)$ & $(123.1890,-110.0440)$ & $(149.5820,-133.6760)$ & $(175.9910,-157.3130)$ \\
\hline
\end{tabular}
\label{field_derivative_zero_table3}
\end{table}

\begin{table}[!htbp]
\centering
\caption{For $\mathcal{A} = 0.002$ and $g = 1.0$}
\begin{tabular}{|c|c|c|c|c|c|}
\hline
& $L=400$ & $L=500$ & $L=700$ & $L=850$ & $L=1000$ \\
\hline
$0$ & $(22.8105,0.0000)$ & $(22.8231,0.0000)$ & $(22.8312,0.0000)$ & $(22.8337,0.0000)$ & $(22.8352,0.0000)$ \\
\hline
$1$ & $(24.4142,-7.8571)$ & $(25.7975,-11.5298)$ & $(30.1132,-19.3048)$ & $(34.1462,-25.0876)$ & $(38.5419,-30.7544)$ \\
\hline
$2$ & $(26.2136,-12.9862)$ & $(29.4579,-18.6441)$ & $(37.2434,-29.3022)$ & $(43.6325,-36.9638)$ & $(50.2515,-44.4609)$ \\
\hline
$3$ & $(27.4552,-19.3792)$ & $(32.5724,-25.8913)$ & $(43.4238,-38.3490)$ & $(51.8345,-47.4332)$ & $(60.3656,-56.4023)$ \\
\hline
$4$ & $(32.0038,-27.6855)$ & $(39.1655,-35.4063)$ & $(53.8027,-50.5481)$ & $(64.9127,-61.7797)$ & $(76.0801,-72.9582)$ \\
\hline
$5$ & $(42.7407,-39.6999)$ & $(53.1177,-49.8987)$ & $(73.9903,-70.1935)$ & $(89.6945,-85.3729)$ & $(105.4200,-100.5350)$ \\
\hline
$6$ & $(70.4395,-62.7598)$ & $(87.9797,-78.4968)$ & $(123.0900,-109.9560)$ & $(149.4360,-133.5450)$ & $(175.7880,-157.1320)$ \\
\hline
\end{tabular}
\label{field_derivative_zero_table4}
\end{table}

\begin{figure}[H]
\centering
\subfloat{\includegraphics[width=.45\linewidth]{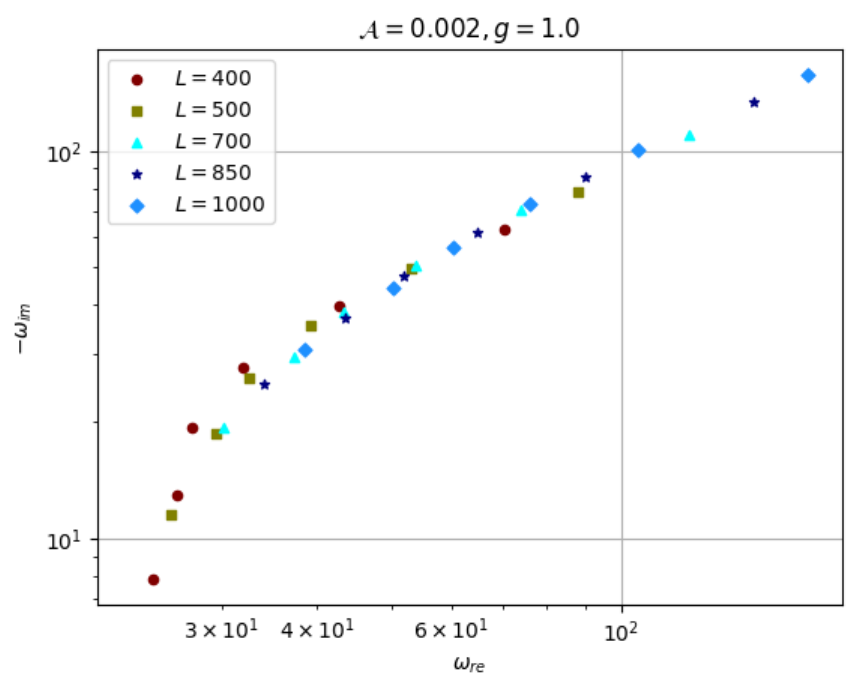}} \hspace{10mm}
\subfloat{\includegraphics[width=.45\linewidth]{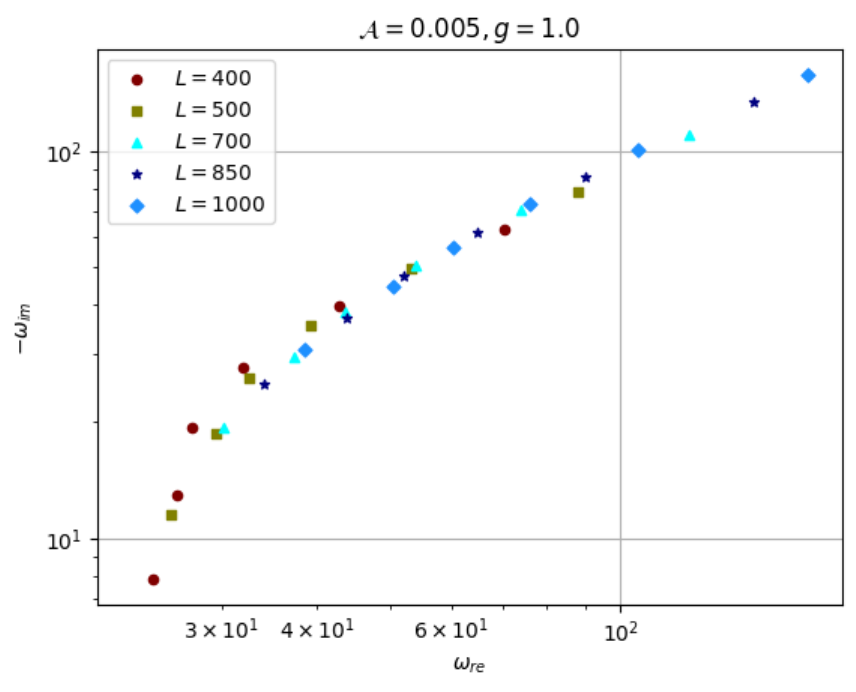}} \par
\subfloat{\includegraphics[width=.45\linewidth]{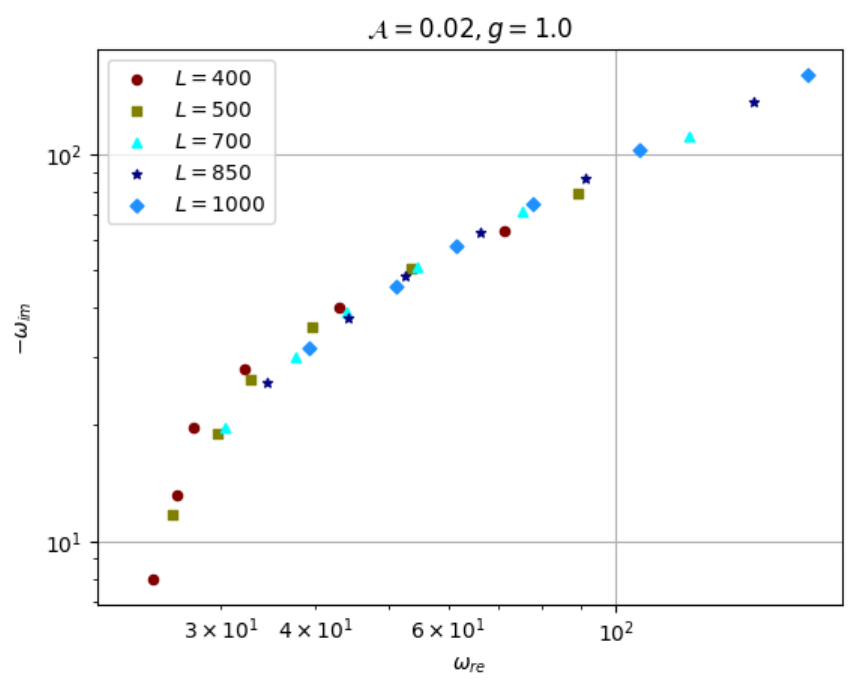}} \hspace{10mm}
\subfloat{\includegraphics[width=.45\linewidth]{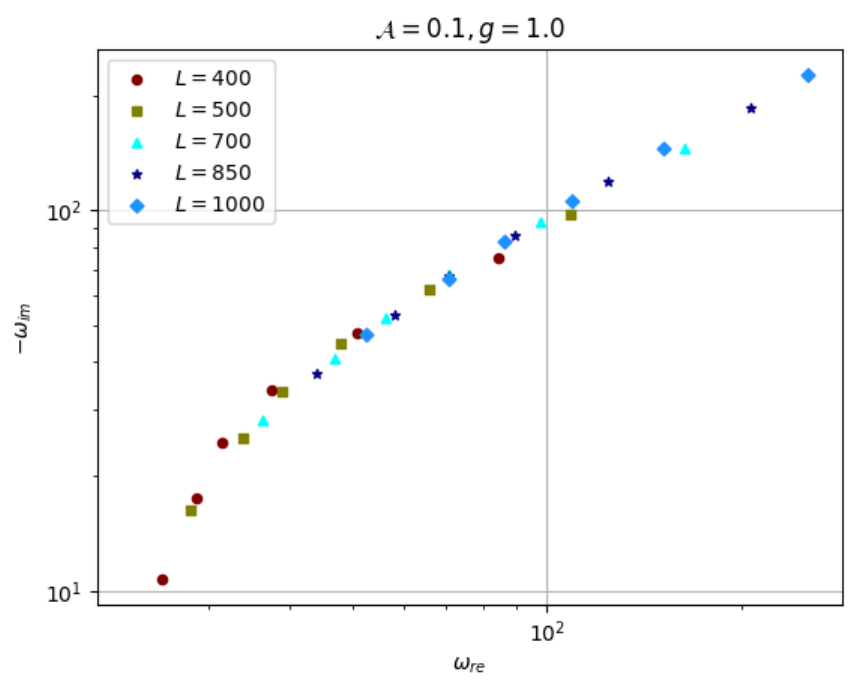}}
\caption{$\Re{(\omega)}$ vs $\Im{(\omega)}$ plot for different modes $(L)$ with different acceleration parameter and a fixed $g=1.0$.}
\label{fig:field_derivative_omega_plot}
\end{figure}

\section{Concluding Remarks}\label{sec:conclusion}
In this paper, we present a thorough analysis of the physical properties of a newly constructed wormhole C--metric, which arises as an axisymmetric solution of Einstein's field equations with a phantom scalar field source. We have argued that the acceleration parameter of the wormhole should be bounded by the amplitude of the potential of the underlying scalar field to make the solution physically viable. Also we have found that the wormhole will not be traversable at all as its throat is located at the origin. Still this spacetime possesses diverse interesting properties. Our investigation reveals intriguing characteristics of this spacetime, including its lack of asymptotic flatness despite the absence of curvature singularities. Furthermore, the study of energy conditions as well as the topological analysis confirms its wormhole nature, emphasizing the importance of distinguishing it observationally from black holes. Notably, the absence of photon orbits poses a challenge for electromagnetic observation, complicating efforts to discern this spacetime from the black holes. But this spacetime can possess the \emph{massive particle orbit}, which may carry the observational interesting features if the detection is carried out using neutrino or alike particles, which is massive as well as can travel long distances without much more interaction with the surroundings. Additionally, we have explored the behavior of quasinormal modes, finding that the space time will result in a continuous spectra if one of the boundary condition is chosen as the field to be outgoing at both infinities. In that case, the field also getting divergent at the throat causing the field to behave in a disconnected manner in two halves of the wormhole. Considering another set of boundary conditions, the radial derivative of effective potential have an asymptotic behavior at the spatial infinities alongwith the finiteness of the field at the throat, we found a continuous spectra of quasinormal modes. We observe a transition from purely oscillatory modes for $n=0$ to damped oscillations with increasing overtone number for $M=0$, which immediately implies that the boundary conditions result in the leakage of field at the infinities for excited modes causing the dissipation in the signal at the detector. Also the relative increment in $\Re{(\omega)}$ and $\Im{(\omega)}$ with the increase in the overtone number $(n)$ indicates that spacetime tries to holds the excitation rather than getting de-excited and coming to the ground state. The reason behind why this phenomena happens and what is its physical interpretation, it requires a thorough analysis of gravitational perturbation of the spacetime. As this is a \emph{non--vacuum axistationary} spacetime, so it is judicious to carry out that gravitational perturbation via Newman--Penrose formalism (citations), which is not only a very recent concept for a non--vacuum spacetime (citations) (gravitational QNM of vacuum black hole C--metric has been studied very recently (citation)), but also it involves many non-intuitive difficulties. Therefore that analysis is left for our future studies. We believe that these intriguing features for this kind of accelerated wormholes spacetimes in a periodic potential will result in phenomenologically and observationally interesting prospects.

\section{Acknowledgement}
SC gratefully acknowledges the support and resources provided by the Presidency University library and online facilities. The authors collectively express their gratitude to the authors of the \texttt{QNMspectral} \cite{QNMspectral_paper} and \texttt{WKB.m} \cite{WKB_paper} packages for making their codes publicly available, which immensely benefited our research efforts. Furthermore, we express our gratitude to Emanuele Berti (JHU) for making his notebooks publicly available. These resources greatly facilitated the dissemination of our research findings.

\appendix
\section{Notations and Conventions}\label{appendix:notations}
Throughout the article, we have chosen the natural unit system $(c=\hbar=8\pi G=1)$ and the mostly positive metric signature $(-,+,+,+)$ has been adopted. The list of symbols and notations used in the article has been given below:
\begin{enumerate}
    \item $\boldsymbol{\sqrt{-\Tilde{g}}~d^4x}$: $4$-dimensional invariant volume element, first appeared in Eq. \eqref{action}.
    \item $\boldsymbol{x}$: spacetime coordinates $(\tau,r,x,\varphi)$ or $(\tau,\rr,x,\varphi)$ or $(\zeta,\rho,x,\varphi)$.
    \item $\Phi(\boldsymbol{x})$: scalar field coupled with, first appeared in Eq. \eqref{action}.
    \item $\mathcal{A}$: rescaled acceleration of wormhole, first appeared in Eq. \eqref{capital_phi}.
    \item $\epsilon$: categorizes the null $(0)$, timelike $(-1)$ and spacelike $(+1)$ geodesics, first appeared in Eq. \eqref{Lagrangian}.
    \item $E$: energy of a particle or photon moving in this spacetime, first appeared in Eq. \eqref{E}.
    \item $\ell$: angular momentum of particle of photon in this spacetime, first appeared in Eq. \eqref{ell}.
    \item $\Bar{\phi}$: test scalar field, first appeared in Eq. \eqref{KleinGordon_equation}.
    \item $\varepsilon$: $g^2-\mathcal{A}^2$, first appeard in Eq. \eqref{rescaled_C-metric2}.  
    \item $\xi$: separation of constant corresponding to angular equation in Klein--Gordon equation, first appeared in Eq. \eqref{angular_equation}.
    \item $L,M$: azimuthal quantum number and magnetic quantum number respectively, associated with spherical harmonics, first appeared in Eq. \eqref{xi_expression}.
    \item $\psi$: radial part of the test scalar field, first appeared in Eq. \eqref{master_equation}.    
\end{enumerate}

\section{Topological properties of the spacetime}\label{topology_wormhole}
By definition \cite{Visser_Lorentzian_wormholes}, in a Lorentzian spacetime, if there exists a compact region denoted by $\Omega$, and if $\Omega$ exhibits a topology resembling $\mathbb{R}\times\Sigma$, where $\Sigma$ represents a three--manifold with non-trivial topology and its boundary $\partial\Sigma$ homeomorphic to $\mathbb{S}^2$, and additionally, if all hypersurfaces $\Sigma$ are spacelike, then it follows that the region $\Omega$ contains a \emph{quasipermanent intra--universe wormhole}. Here, ``quasipermanent'' refers that the wormhole exists for a finite non--zero duration of time.

For the case of line element \eqref{C-metric}, the topology of spacelike hypersurface $\Sigma$ can be determined by the constant $\varphi$ hypersurfaces (let, $\Sigma'$) due to the symmetry. Then the induced metric over $\Sigma'$ can be written as
\begin{equation}
    ds_{\Sigma'}^2 = \dfrac{1}{(1+\A\rr x)^2} \left[\dfrac{1+\A^2x^2}{1+g^2+(g^2-\A^2)\rr^2} d\rr^2 + \dfrac{1+\rr^2}{1-x^2} dx^2\right] \equiv h_{\mu\nu} dx^\mu dx^\nu ~ \text{(let)}
\end{equation}
where $\Omega$ to $\Sigma'$ and $\Sigma$ to $\Sigma'$ projectors are essentially the Kronecker deltas and the normal to $\Sigma'$ is also the same. The triviality or non-triviality of topology of the manifold $\Sigma'$ can now be determined by its Euler characteristic $(\chi)$ from the \emph{Gauss--Bonnet formula}. If the Gaussian curvature of this $2$ dimensional manifold is being denoted by $K_G$, then by \emph{Gauss's Theorema Egregium} \cite{Gaussian_Curvature__book}, we have
\begin{equation}
    K_G = -\dfrac{1}{2\sqrt{h_{11}h_{22}}} \left[\partial_x \left(\dfrac{\partial_x h_{11}}{\sqrt{h_{11}h_{22}}}\right) + \partial_\rr \left(\dfrac{\partial_\rr h_{22}}{\sqrt{h_{11}h_{22}}}\right)\right] \label{Gaussian_Curvature_expression}
\end{equation}
Considering $\Sigma'$ to be orientable compact $2$ dimensional Riemannian manifold, Gauss--Bonnet theorem reads
\begin{equation}
    \chi = \dfrac{1}{2\pi} \displaystyle\int_{\Sigma'} K_G \sqrt{h}~d\rr~dx + \dfrac{1}{2\pi} \displaystyle\int_{\partial\Sigma'} k_g d\mathcalligra{s} \label{Gauss_Bonnet_theorem}
\end{equation}
where $k_g$ is the geodesic curvature of the boundary $\partial\Sigma'$ and $d\mathcalligra{s}$ be its line element. In this case, $k_g=0$. Therefore using Eq. \eqref{Gaussian_Curvature_expression}, rest of the part of Eq. \eqref{Gauss_Bonnet_theorem} reduces to
\begin{equation}
    \chi = -\dfrac{1}{4\pi} \left[\displaystyle\int_{\Sigma'}  d\rr \left.\dfrac{\partial_x h_{11}}{\sqrt{h_{11}h_{22}}}\right|_{x\to1} - \displaystyle\int_{\Sigma'}  d\rr \left.\dfrac{\partial_x h_{11}}{\sqrt{h_{11}h_{22}}}\right|_{x\to-1} + \displaystyle\int_{\Sigma'} dx \left.\dfrac{\partial_\rr h_{22}}{\sqrt{h_{11}h_{22}}}\right|_{r\to\infty} - \displaystyle\int_{\Sigma'} dx \left.\dfrac{\partial_\rr h_{22}}{\sqrt{h_{11}h_{22}}}\right|_{r\to-\infty}\right] \label{EulerCharacteristic_final_expression}
\end{equation}
All the limiting terms in Eq. \eqref{EulerCharacteristic_final_expression} identically vanishes resulting $\chi=0$, resulting this to be a quasipermanent intra–universe wormhole spacetime.

\section{Perturbation series analysis of scalar field angular equation}\label{perturbative_series_angular}
In the zero acceleration limit, the angular equation \eqref{angular_equation} of the scalar perturbation can be written as
\begin{equation}
    \Hat{\mathbb{L}}(x) \Theta_0(x) + \left[\xi_0 - \dfrac{1}{3} - \dfrac{M^2}{1-x^2}\right] \Theta_0(x) = 0 \label{zero_acceleration_angular_appendix}
\end{equation}
where $ \Hat{\mathbb{L}}$ is a Strum--Liouville operator defined in range $x\in[-1,+1]$ as
\begin{equation}
    \Hat{\mathbb{L}}(x) \equiv  \left(1-x^2\right) \dfrac{d^2}{dx^2} - 2x \dfrac{d}{dx}
\end{equation}
This is self-adjoint as well as Hermitian operator i.e. it has complete set of orthonormal eigenvectors alongwith real eigenvalues. Eq. \eqref{zero_acceleration_angular_appendix} can readily be solved by means of associated Legendre polynomials $P_{LM}(x)$, which will be finite over the entire range and to guarantee the necessary convergence of the associated Frobenius series, it needs $\xi_0=L(L+1)+\frac{1}{3}$. This is the zeroth order approximation of the eigenvalue and the eigenvector. For the higher order corrections, the eigenvalue $\xi$ and the eigenfunction $\Theta$ can be expanded in a perturbation series as
\begin{eqnarray}
    \xi &=& \xi_0 + \A^2\xi_2 + \A^4\Theta_4(x) + \mathcal{O}(\A^6) \label{xi_perturbation_series} \\
    \Theta(x) &=& \Theta_0(x) + \A^2\Theta_2(x) + \A^4\Theta_4(x) + \mathcal{O}(\A^6) \label{Theta_perturbation_series}
\end{eqnarray}
where $\Theta_0(x) = P_{LM}(x)$. Substituting Eqs. \eqref{xi_perturbation_series} and \eqref{Theta_perturbation_series} in Eq. \eqref{angular_equation} and expanding the terms in Taylor series, we have after comparing the coefficients of $\A^2$ on both sides, 
\begin{equation}
    \left[\Hat{\mathbb{L}}(x) + \xi_0 - \dfrac{1}{3}\right]\Theta_1(x) + \xi_2 \Theta_0(x) = 0 \label{angular2_appendix}
\end{equation}
Multiplying both sides of Eq. \eqref{angular2_appendix} by $\Theta_0^*(x)$ (complex conjugated function of $\Theta_0(x)$) and using the fact that $\Hat{\mathbb{L}}(x)$ is a self-adjoint operator being its Strum--Liouville nature, we arrive at Eq. \eqref{xi_expression}. This is the $2^{\text{nd}}$ order perturbative approximation of the eigenvalue, where the $1^{\text{st}}$ and other odd order corrections are zero.

\section{Heun function}\label{Heun_function}
The general Heun equation is given by the following equation \cite{Heun_original_paper}:
\begin{equation}
    \dfrac{d^2u}{d\Tilde{z}^2} + \left(\dfrac{\Tilde{\gamma}}{\Tilde{z}} +\dfrac{\Tilde{\delta}}{\Tilde{z}-1}+\dfrac{\Tilde{\epsilon}}{\Tilde{a}}\right) \dfrac{du}{d\Tilde{z}} + \dfrac{\Tilde{\alpha}\Tilde{\beta}\Tilde{z}-\Tilde{q}}{\Tilde{z}(\Tilde{z}-1)(\Tilde{z}-\Tilde{a})} u = 0
\end{equation}
where the parameters satisfy the Fuchsian relation \(\Tilde{\epsilon} = 1+\Tilde{\alpha}+\Tilde{\beta}-\Tilde{\gamma}-\Tilde{\delta}\). This equation has four regular singular points at $0,1,\Tilde{a},\infty$ and \textit{any} differential equation with four regular singular points (such as Hypergeometric differential equations or Lamé equation) can be transformed to this equation by a change of variables \cite{Ronveaux_book}. For each singular points, it is possible to construct Frobenius series solution, but one cannot get a recursion relation between two consecutive coefficients. We have a relation at least between three coefficients \cite{Heun_original_paper}. The solutions are convergent inside a circle, whose radius is generically determined by the distance from the nearest singularity.

It turns out that for infinite set of values of the parameter $\Tilde{q}$, there are solutions which are analytic at $0$ and at $1$. These are called Heun functions, whereas those which are analytic only at one point are called \textit{local Heun functions}, denoted by \(\Hl(\Tilde{a},\Tilde{q};\Tilde{\alpha},\Tilde{\beta},\Tilde{\gamma},\Tilde{\delta};\Tilde{z})\). Local Heun function that is also regular at $\Tilde{z}=1$ is called Heun function, denoted by \(\Hf(\Tilde{a},\Tilde{q};\Tilde{\alpha},\Tilde{\beta},\Tilde{\gamma},\Tilde{\delta};\Tilde{z})\). For integer values of one of $\Tilde{\alpha}$, $\Tilde{\gamma}-\Tilde{\alpha}$, $\Tilde{\delta}-\Tilde{\alpha}$, $\Tilde{\epsilon}-\Tilde{\alpha}$, and for special finite values of $\Tilde{q}$, solutions analytic at three singularities exist, the so-called \textit{Heun polynomials}, denoted by \(\Hp(\Tilde{a},\Tilde{q};\Tilde{\alpha},\Tilde{\beta},\Tilde{\gamma},\Tilde{\delta};\Tilde{z})\) \cite{Heun_function_applications}. So if a Local Heun function is tried to make regular at $\Tilde{z}=1$, then the accessory parameter $\Tilde{q}$ have to be quantized \cite{Sleeman_Heun@z1} and in that case, it should satisfy Eq. \eqref{Sleeman--Meixner--Schafke_theorem}.

The local Heun function can be normalized to unity at $\Tilde{z}=0$ \cite{Kerr_dS_Heun} as
\begin{equation}
\Hl(\Tilde{a},\Tilde{q};\Tilde{\alpha},\Tilde{\beta},\Tilde{\gamma},\Tilde{\delta};0) = 1
\end{equation}
and its derivative with respect to $\Tilde{z}$ at $\Tilde{z}=0$ takes the value $\frac{\Tilde{q}}{\Tilde{a}\Tilde{\gamma}}$ and diverges at $\Tilde{z}=1$.

In Mathematica \cite{Mathematica}, the local Heun function and its first derivative are implemented by the library functions as \texttt{HeunG$[\Tilde{a},\Tilde{q},\Tilde{\alpha},\Tilde{\beta},\Tilde{\gamma},\Tilde{\delta},\Tilde{z}]$} and \texttt{HeunGPrime$[\Tilde{a},\Tilde{q},\Tilde{\alpha},\Tilde{\beta},\Tilde{\gamma},\Tilde{\delta},\Tilde{z}]$} respectively. These have the same normalization and other properties discussed above. Globally \texttt{HeunG} gives analytic continuation of $\Hl$. Though sometimes multi-valuedness causes a problem in numerical computations, all our computations are done inside the convergence circle. Hence there is no multi-value problem.

\medskip
\nocite{*}
\bibliography{main}
\end{document}